\documentclass[fleqn,11pt,twocolumn,]{wlscirep}
\usepackage[utf8]{inputenc}
\usepackage[T1]{fontenc}
\usepackage{mathtools,multirow,amssymb,commath,siunitx,gensymb}
\usepackage{xcolor,colortbl}
\usepackage{pdflscape} 
\def\fillandplacepagenumber{%
 \par\pagestyle{empty}%
 \vbox to 0pt{\vss}\vfill
 \vbox to 0pt{\baselineskip0pt
   \hbox to\linewidth{\hss}%
   \baselineskip\footskip
   \hbox to\linewidth{%
     \hfil\thepage\hfil}\vss}}
\usepackage{changepage}
\title{Machine-learning potentials for structurally and chemically complex MAB phases: strain hardening and ripplocation-mediated plasticity}

\author[1,2,*]{Nikola Koutn\'a}
\author[1,2]{Shuyao Lin}
\author[2,3]{Lars Hultman}
\author[2]{Davide G. Sangiovanni}
\author[1]{Paul H. Mayrhofer}

\affil[1]{Institute of Materials Science and Technology, TU Wien, Getreidemarkt 9, A-1060 Vienna, Austria}

\affil[2]{Department of Physics, Chemistry, and Biology (IFM), Link\"{o}ping University, SE-58183 Link\"{o}ping, Sweden}

\affil[3]{Center for Plasma and Thin Film Technologies, Ming Chi University of Technology, New Taipei City 24301, Taiwan}

\affil[*]{nikola.koutna@tuwien.ac.at}

\begin{abstract}
Though offering unprecedented pathways to molecular dynamics (MD) simulations of technologically-relevant materials and conditions, machine-learning interatomic potentials (MLIPs) are typically trained for ``simple'' materials and properties with minor size effects. 
Our study of MAB phases (MABs)---alternating transition metal boride (MB) and group A element layers---exemplifies that MLIPs for complex materials can be fitted and used in a high-throughput fashion: for predicting structural and mechanical properties across a large chemical/phase/temperature space.
Considering group 4--6 transition metal based MABs, with $\text{A}=\text{Al}$ and the 222, 212, and 314 type phases, three MLIPs are trained and tested, including lattice and elastic constants calculations at temperatures $T\in\{0,300,1200\}$~K, extrapolation grade and energy (force, stress) error analysis for $\approx{3\cdot10^6}$ {\it{ab initio}} MD snapshots. 
Subsequently, nanoscale tensile tests serve to quantify upper limits of strength and toughness attainable in single-crystal MABs at 300~K as well as their temperature evolution.
In-plane tensile deformation is characterised by relatively high strength, \{110\}$\langle001\rangle$ type slipping, and failure by shear banding. 
The response to [001] loading is softer, triggers work hardening, and failure by kinking and layer delamination. 
Furthermore, W$_2$AlB$_2$ able to retard fracture via ripplocations and twinning from 300 up to 1200~K.
\end{abstract}

\begin{document}

\flushbottom
\maketitle

\thispagestyle{empty}



\section{Introduction}
The swift development in the field of machine-learning interatomic potentials~\cite{mishin2021machine} (MLIPs) opens the door to molecular dynamics simulations (MD) of complex materials under experiment-relevant conditions, reaching near {\it{ab initio}} accuracy at greatly reduced computational costs~\cite{smith2017ani,shapeev2020elinvar}.
MLIPs are non-empirical models that learn atomic properties from a reference {\it{ab initio}} dataset, using mathematically convenient and possibly physically-motivated descriptors (e.g., moment tensors~\cite{MTP}, atomic clusters~\cite{drautz2019atomic}) to encode atomic environments.
MLIPs are improvable by active learning exploiting certain extrapolation control (e.g., the MaxVol grade via the D-optimality criterion~\cite{MLIP}). 
However, fundamental challenges---such as (i) construction of robust and transferable training sets, and (ii) validation of MLIP-MD simulations beyond {\it{ab initio}} length scales---make efficient MLIP training for structurally and/or chemically complex materials and defects difficult.
Especially MLIPs for simulations including wide range of mechanical strains and strain-activated defects are non-trivial, and we attempt making a step forward in this direction.

Already MLIP development for elastic constants 
calculations is far from routine~\cite{tasnadi2021efficient,gubaev2021finite}.  
Modelling nucleation and dynamics of defects governing macroscopic mechanical behaviour poses even bigger challenges~\cite{mortazavi2023atomistic,deng2024room}, as extended defects are typically too large to be explicitly labelled by {\it{ab initio}} calculations and included in the training database.
Examples of MLIPs tailor-trained for accurate simulations of dislocations~\cite{zhang2024efficiency} and grain boundaries~\cite{nishiyama2020application,ito2024machine} most often concern elemental metals. 
MLIP-MD mechanistic simulations at the nanoscale have been sparse; including tensile testing and fracture toughness predictions for C-based materials~\cite{mortazavi2022low,mortazavi2021first,mortazavi2023atomistic,qamar2023atomic}, BaZrO$_3$~\cite{wang2023machine}, TiB$_2$~\cite{lin2024machine}, as well as nanoindentation simulations for W~\cite{dominguez2023nanoindentation}, Si, AlN, SiC, and BC$_2$N~\cite{podryabinkin2022nanohardness}.
Routine predictions of strength and toughness related quantities and mechanisms, however, have been hindered by the lack of MLIPs for structurally complex materials (e.g., multicomponent alloys, superlattices) and/or their unproved transferability (especially in case of universal pre-trained potentials such as M3GNET~\cite{chen2022universal}, CHGNET~\cite{deng2023chgnet}, or MACE~\cite{batatia2023foundation}).

MAB phases~\cite{kota2020progress} (MABs) alternate atomically-thin layers of ceramic-like transition metal boride, M--B, with mono- or bi-layers of an A-element (A$=$Al, Si, Ga, In,\ldots).
The modulated charge density distribution, with predominantly covalent--ionic bonds separated by metallic bonds, renders a combination of properties typical for ceramics (e.g., high-temperature strength and oxidation resistance) as well as for metals (e.g., high fracture toughness and damage tolerance).
Thus, MABs are attractive for high-temperature structural applications, as protective coatings, sensors, low friction surfaces, electrical contacts, tunable damping films for microelectromechanical systems, or as a basis of 2D materials for new-generation nanodevices~\cite{dahlqvist2020theoretical,wang2019discovery,chakraborty2018soft,zhou2021boridene}. 
With typical formula M$_{n+1}$AB$_{2n}$ ($n=1,2,3,\ldots$),
MABs are rich in structures, chemistry, and bonding motifs; including five experimentally known hexagonal or orthorhombic phase prototypes: MAB (222 type, $Cmcm$~\cite{zhang2023experimental}), M$_2$AB$_2$ (212 type,  $Cmmm$~\cite{kota2018synthesis} and $P\overline{6}m2$~\cite{wang2019discovery}),  M$_3$AB$_4$ (314 type, $Pmmm$~\cite{kota2018magnetic}), M$_4$AB$_6$ (416 type, $Cmmm$~\cite{ade2015ternary}), and other stable phases predicted by {\it{ab initio}} calculations~\cite{carlsson2022theoretical,koutna2024phase,khazaei2019novel}. 
The vast chemical and phase space together with numerous competing phases makes systematic screening of this material class difficult and results in relatively few MABs synthesised till today (Ti$_2$InB$_2$~\cite{wang2019discovery}, Cr$_2$AlB$_2$~\cite{berastegui2020magnetron,kota2018magnetic,ade2015ternary}, Cr$_3$AlB$_4$~\cite{kota2018magnetic,ade2015ternary}, Cr$_4$AlB$_6$~\cite{ade2015ternary}, MoAlB~\cite{zhang2023experimental,chen2019compressive,achenbach2019synthesis,evertz2021low,sahu2022defects}, WAlB~\cite{zhang2023experimental,roy2023low}, Fe$_2$AlB$_2$~\cite{liu2018rapid}, Mn$_2$AlB$_2$~\cite{kota2018synthesis}).

Mechanical response of MABs has been investigated by 0~K density functional theory (DFT) calculations including phenomenological descriptors of  strength and ductility~\cite{liu2020new,dai2018first,lind2021plane,koutna2024phase}, toughness and damage tolerance~\cite{bai2017density,bai2019phase}. 
Experimental studies concerned indentation hardness (e.g., MoAlB~\cite{achenbach2019synthesis,evertz2021low}, Mn$_2$AlB$_2$~\cite{kota2018synthesis}) and fracture toughness measurements (Fe$_2$AlB$_2$~\cite{li2017rapid}), as well as crack deflection and crack healing (MoAlB~\cite{lu2019crack,lu2019thermal}, Fe$_2$AlB$_2$~\cite{bai2019high}).
Currently lacking atomistic simulations could help understanding how mechanical properties and atomic-scale deformation mechanisms of MABs depend on their chemistry and structure.
They could also provide hints for a very fundamental question regarding deformation mechanisms; namely, whether MABs can nucleate so-called {\it{ripplocations}}~\cite{barsoum2020ripplocations}.
Ripplocations are atomic layer ripples claimed to be the universal deformation mechanism in layered materials upon edge-on basal plane compression, observed at nm-to-km length scales (in MAX phases as well as geological formations)~\cite{plummer2021origin}.
Despite the laminated character of MABs and their similarity to MAX phases, no ripplocations have been reported, Chen {\it{et al.}}~\cite{chen2019compressive} even suggested that MABs {\it{do not}} ripplocate.

In this study, we present a systematic screening of 222, 212, and 314 type MABs containing group 4--6 transition metals, M$=$(Ti, Ta, W), and A$=$Al, focusing on structural parameters, mechanical strength, toughness, and fracture mechanisms as a function of temperature.
The chosen MABs have been shown dynamically stable~\cite{koutna2024phase}, some are isoelectronic to already synthesised MABs~\cite{berastegui2020magnetron,kota2018magnetic,ade2015ternary}, and WAlB is experimentally known~\cite{zhang2023experimental,roy2023low}. 
Our results are derived from MD simulations with own MLIPs, validated in terms of lattice and elastic constants at $T\in\{0,300,1200\}$~K, as well as via extrapolation grade and energy (force, stress) analysis against $\approx{3\cdot10^6}$ {\it{ab initio}} MD snapshots from finite-temperature mechanical tests.
Uniaxial tensile simulations at the nanoscale ($\approx{300,000}$-atom supercells) serve for quantitative predictions of theoretical strength and toughness as a function of $T$.
Furthermore, we elucidate deformation mechanisms---including ripplocations---and crack nucleation patters characteristic for group 4--6 transition metal based MABs.


\section{Methods}

\subsection{Ab initio dataset}\label{Sec: 2.1}
Zero Kelvin {\it{ab initio}} and finite-temperature Born-Oppenheimer {\it{ab initio}} MD calculations were performed employing the VASP~\cite{VASP-1} code together with the projector augmented wave (PAW)~\cite{VASP-2} method and the Perdew-Burke-Ernzerhof exchange-correlation functional~\cite{PBEsol}.
The plane-wave cut-off energy was 300~eV, the reciprocal space was sampled with a single $\Gamma$-point, and the time step was always 1~fs.
Considering M$=$(Ti, Ta, W) and A$=$Al, models of the 222 (M$_2$A$_2$B$_2$ or simply MAB), 212 (M$_2$AB$_2$), and 314 (M$_3$AB$_4$) type MABs were based on the orthorhombic Pmm2, Cmcm, and Pmmm structures~\cite{kota2020progress}, with the $x\parallel[100]$, $y\parallel[010]$, and $z\parallel[001]$. 
Calculations for 222, 212, and 314 type MABs were performed in 864-, 720-, 576-atom supercells (6$\times$6$\times$2), respectively, equilibrated at $T\in\{300,1200\}$~K during (i) an isobaric-isothermal (NPT) simulation with Parrinello-Rahman barostat~\cite{NPT} and Langevin thermostat (with a 1~fs timestep, for at least 6~ps), followed by (ii) a simulation with the canonical (NVT) ensemble and the Nosé-Hoover thermostat (with a 1~fs timestep, for 2~ps), using time-averaged lattice parameters from the last two 2~ps of step (i).

To generate training and validation data for MLIPs (targeted to finite-temperature tensile simulations for single-crystal MABs), series of tensile tests were performed for each phase prototype and elemental combination at 300 and 1200~K.
Specifically, uniaxial strain was applied in the [001], [010], and [001] direction, respectively, with a 2\% strain step (including 3~ps NVT relaxation) until material's fracture. 
The simulations followed Refs.~\cite{Method1,Method2,koutna2022atomistic}.
Additionally, shearing along the [100](010) and [100](001) slip systems and volumetric compression were simulated at 1200~K, always starting from equilibrated structures.
Shearing followed the setup of the above tensile tests, whereas volumetric compression was modelled via shrinking lattice parameters ($a$, $b$, $c$) by up to 5\% and maintaining for 3~ps under the NVT ensemble.
Furthermore, shock uniaxial tension was simulated by elongating the $a$ ($b$, $c$) lattice parameter by 125--150\%, initializing atoms at ideal lattice sites, and equilibrating for 3~ps under the NVT ensemble. 
Environments produced by such simulations were suggested useful for describing the onset of fracture during nanoscale tensile tests~\cite{lin2024machine}.

\subsection{Machine-learning interatomic potentials (MLIPs)}\label{Sec: 2.2}
All MLIPs were trained in the moment tensor potential (MTP~\cite{shapeev2016moment}) formalism using the {\texttt{mlip-2}}~\cite{MLIP} package with 16g MTPs, 5.5~\AA\ cut-off radius, 1500 iterations of the Broyden-Fletcher-Goldfarb-Shanno algorithm~\cite{fletcher2013practical}, and $\{1;0.01;0.01\}$ weights for energies, forces, stresses.
MLIP(Ti-MABs), MLIP(Ta-MABs), and MLIP(W-MABs) were trained in an iterative active learning procedure, consistently with our previous work~\cite{lin2024machine}.
The split of {\it{ab initio}} MD data between randomly initialised training set (TS), learning set (LS), and validation set (VS) was 0.5\%--79.5\%--20\%.
Series of intermediate MLIPs were trained on configurations from a particular mechanical test, e.g., [010]-tensile loading of Ta$_2$AlB$_2$, and gradually merged/up-fitted until providing {\it{accurate}} extrapolation on all configurations in the LS, as indicated by the extrapolation grade ($\gamma<2$; consistent with Refs.~\cite{lin2024machine,podryabinkin2023mlip,shapeev2020elinvar}). 
 
\subsection{Molecular dynamics simulations}
MD simulations employed the LAMMPS code~\cite{LAMMPS} interfaced with the {\texttt{mlip-2}} package~\cite{MLIP}.
Computational setups consistent with the above described {\it{ab initio}} MD calculations was used for validation.
Molecular statics was performed when comparing against 0~K {\it{ab initio}} data.
Finite-temperature tensile test were conducted in $\approx{300,000}$-atom supercells with dimension of 15$^3$~nm$^3$, and $x\parallel[100]$, $y\parallel[010]$, $z\parallel[001]$ (convergence of lattice parameters and stress/strain curves was tested in 10$^3$ to 10$^5$-atom supercells).
All MABs were equilibrated for $20$\;ps (1~fs time step) at the targeted temperature under the isobaric-isothermal (NPT) ensemble coupled to the Nosé-Hoover thermostat.
Uniaxial tensile strain was applied along the [100], [010] or [001] direction, respectively, with a constant strain rate (50~\AA/s), 
accounting for the Poisson's contraction (via NPT).

\section{Results and discussion}

\subsection{Atomic-scale simulations} \label{Sec: validation}
To screen structural and mechanical properties of group 4--6 transition metal based MABs---with M$=$(Ti, Ta, W), A$=$Al
, and experimentally known orthorhombic 222, 212, and 314 type structures---we developed three MLIPs: MLIP(Ti-MABs), MLIP(Ta-MABs), and MLIP(W-MABs). 
First, they are employed for atomic-scale simulations concerning lattice and elastic constants as well as various mechanical tests, and validated against DFT/{\it{ab initio}} MD data at temperatures $T\in\{0,300,1200\}$~K.  

\begin{figure}[h!t!]
    \centering
    \includegraphics[width=1\columnwidth]{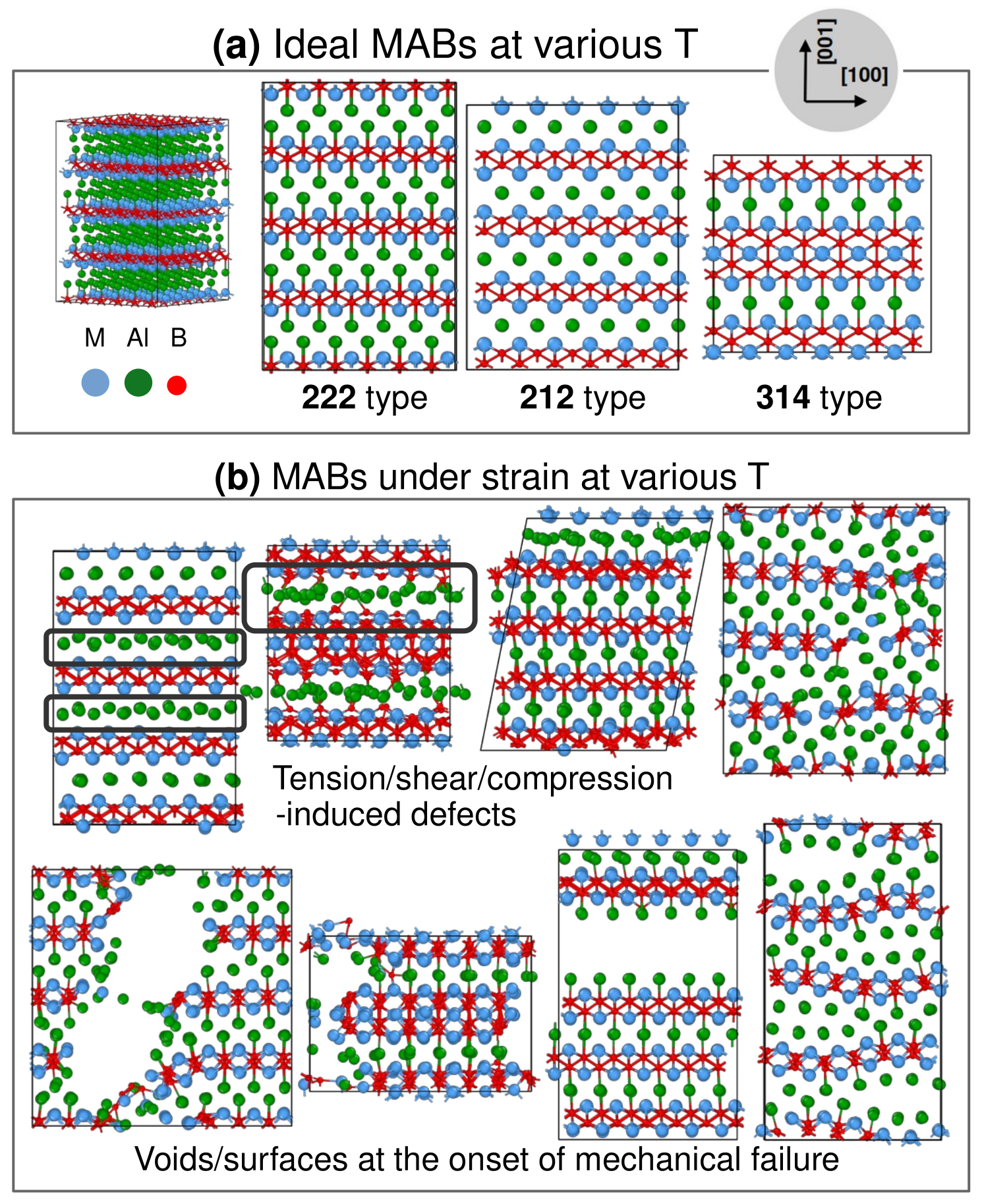}
    \caption{
    \small
    {\bf{Illustration of atomic environments present in training sets of our MLIPs}}. 
    (a) Ideal defect-free MABs with the 222, 212, and 314 type structure. 
    The blue, green, and red spheres represent M$=$(Ti, Ta, W), A$=$Al, and boron, respectively.
    (b) MABs under various strain conditions, as extracted from finite-temperature {\it{ab initio}} MD mechanical tests. 
    All supercells have $\approx{10^3}$ atoms ($\approx{2^3}$~nm$^3$). 
    }
\label{FIG: atomic environments}
\end{figure}

Fig.~\ref{FIG: atomic environments} exemplifies atomic environments in our training sets, including stacking faults, distorted and/or intermixed layers, voids, and surfaces. 
These result from uniaxial/volumetric/shear strain in finite-temperature {\it{ab initio}} MD, i.e., were not hand-crafted using any material-specific knowledge.
Consistently with the underlying {\it{ab initio}} calculations, training and validation errors of energies, forces, and stresses, quantified by the mean square error (RMSE), are below 3~meV/at., 0.14~eV/\AA, and 0.2~GPa for MLIP(Ti-MABs); 20~meV/at., 0.20~eV/\AA, and 0.3~GPa for MLIP(Ta-MABs); and 9~meV/at., 0.23~eV/\AA, and 0.7~GPa for MLIP(W-MABs).
Note that our validation set did not contain binary or other non-MAB compounds, as transferability to these was not targeted and up-fitting using own {\it{ab initio}} MD dataset for TiB$_2$ (TaB$_2$, WB$_2$) has slightly decreased accuracy for MABs.

In line with previous DFT phonon calculations~\cite{koutna2024phase}, both {\it{ab initio}} MD and MLIP-MD indicate dynamical stability of the nine studied MABs (as time-averaged atomic vibrations in $\approx{10^3}$-atom supercells, at $T\in\{300, 1200\}$~K, remain near ideal sites).
Thus, besides the experimentally known WAlB~\cite{zhang2023experimental,roy2023low}, also the other systems are at least metastable, though their formation may strongly depend on competing phases~\cite{khazaei2019novel} and kinetics of the synthesis process.

Tab.~\ref{TAB: aLat, Cij} lists structural and elastic properties predicted at $T\in\{0,300,1200\}$~K.
Differences between MLIP MD and DFT/{\it{ab initio}} MD in terms of lattice parameters and equilibrium volumes are small, reaching ($0.19\pm0.2$)\%.
MLIP predictions may be in fact more accurate as the corresponding (time-averaged) stress tensor components are closer to zero, i.e., to the ideal equilibrium at given $T$. 
Irrespective of M element, the 314 phase exhibits the lowest volume, followed by the 212 and 222 type MABs.
Note that the 314, 212, and 222 phase were indicated as the energetically most favourable MABs for M$=$Ti, M$=$Ta, and M$=$W, respectively~\cite{koutna2024phase}.

\begin{figure*}[h!t!]
    \centering
    \includegraphics[width=2\columnwidth]{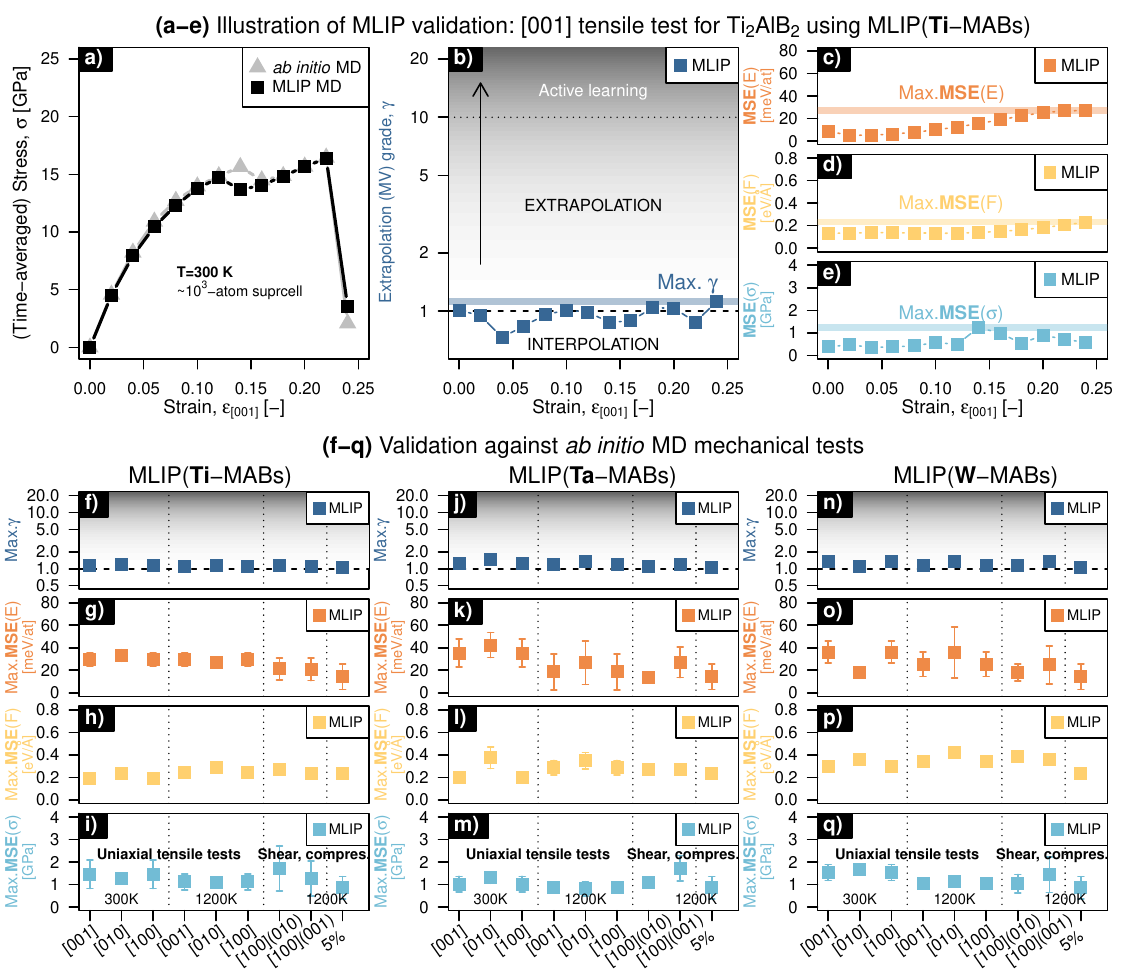}
\caption{ \footnotesize
{\bf{Validation of here-developed MLIPs against {\it{ab initio}} MD mechanical tests}}.
Total of 54 atomic-scale mechanical tests (see the Methods) was simulated for the nine studied MABs with the orthorhombic 222, 212, and 314 type structures (listed in Tab.~\ref{TAB: aLat, Cij}).
(a) Comparison of MLIP-MD and {\it{ab initio}} MD stress/strain curves for Ti$_2$AlB$_2$ during [001] tensile loading at 300~K. 
(b) Evolution of the corresponding extrapolation grades, $\gamma$, where $\gamma\leq1$ ($\gamma>1$) indicates interpolation (extrapolation). 
Residual mean square errors, RMSEs, of MLIP-predicted (a) energies, $E$; (b) forces, $F$; and (c) stresses, $\sigma$; evaluated for all configurations from {\it{ab initio}} MD (2700 configurations at each strain step).  
The horizontal lines in (b--e) guide the eye for the maximum $\gamma$, Max.$\gamma$; and the maximum RMSEs, Max.RMSE($E$), Max.RMSE($F$), and Max.RMSE($\sigma$); reached during the tensile test. 
Analogically obtained Max.$\gamma$, Max.RMSE($E$), Max.RMSE($F$), and Max.RMSE($\sigma$) are shown for various mechanical tests---uniaxial [001], [010], and [100] tensile deformation at 300 and 1200~K; [100](010) and [100](001) shear deformation at 1200~K; and 5\% volumetric compression at 1200~K---for all (f-i) Ti-based, (j--m) Ta-based, and (n--q) W-based MABs.
Note that data in panels (f--q) is derived from validation against $\approx{3\cdot10^6}$ snapshots from {\it{ab initio}} MD calculations. 
} 
\label{FIG: validation}
\end{figure*}

Furthermore, Tab.~\ref{TAB: aLat, Cij} shows temperature evolution of elastic constants, $C_{ij}$ (computed from second-order polynomial fits of stress/strain data from tensile and shear tests).
The comparison of MLIP-predicted $C_{11}$, $C_{22}$, and $C_{33}$---approximating directional Young's moduli $E_{[100]}$, $E_{[010]}$, and $E_{[001]}$---with own DFT/{\it{ab initio}} MD values yields ($7\pm5$)\% difference.
Calculations of the full elastic tensor (9 independent $C_{ij}$s \cite{mouhat2014necessary}) for selected MABs and temperatures indicate similar agreement for the remaining $C_{ij}$s. 
While MLIPs trained particularly for $C_{ij}$ predictions can achieve higher accuracy (e.g., MLIPs for $\beta$-TiNbZr~\cite{mukhamedov2024machine}, bcc-TiZrHfTa~\cite{gubaev2021finite}, or fcc-TiAlN~\cite{tasnadi2021efficient}), presumably more transferable MLIPs---applicable to simulations with high strains and/or multiple phase structures---yield differences from DFT/{\it{ab initio}} MD similar to ours, e.g., $\Delta_{\text{DFT}}(C_{ij})\approx{5}\%$ for TiB$_2$~\cite{lin2024machine} or $\Delta_{\text{DFT}}(C_{ij})=0$--25~\% for W~\cite{xie2023ultra}.
The obtained differences from corresponding DFT/{\it{ab initio}} MD values are also well within the range of differences between 0~K $C_{ij}$s for the same MAB phase from various DFT studies (c.f. Refs.~\cite{koutna2024phase,bai2017density,liu2020new} for MoAlB).

As expected, $C_{ij}$s of all MABs soften with $T$. 
Irrespective of $T$, all MABs are predicted to be elastically stiffer in-plane, parallel to M--B/A layers, compared to out-of-plane, orthogonal to M--B/A layers.
With some exceptions (e.g., $C_{22}$ of Ta-MABs), $C_{11}$, $C_{22}$, and $C_{33}$ increase when changing 222$\to$212$\to$314 type phase, indicating increasing resistance to strains along main crystallographic directions.

Considering targeted predictions of theoretical tensile strength, toughness, and deformation mechanisms, our MLIPs are further validated against snapshots from finite-temperature mechanical tests, comprising total of $\approx{3\cdot10^6}$ {\it{ab initio}} MD configurations (Fig.~\ref{FIG: validation}).
In particular, (i) uniaxial [001], [010], and [100] tensile deformation, (ii) [100](010) and [100](001) shear deformation, and (iii) volumetric compression are simulated at 300~K and/or at 1200~K for each MAB phase.
As supercells sizes affordable in {\it{ab initio}} MD (here $\approx{10}^3$-atom supercells; $\approx{2^3}$nm$^3$) severely constrain nucleation of extended defects, we focus on showing consistency between MLIP and {\it{ab initio}} MD data, omitting discussion of the predicted mechanical properties.

Quantitative agreement between MLIP and {\it{ab initio}} MD mechanistic simulations is shown by closely overlapping stress/strain curves (Fig.~\ref{FIG: validation}a: example of room-temperature tensile test for Ti$_2$AlB$_2$) and the same/similar stress release mechanisms (e.g., layer bending and delamination).
Note that due to finite-temperature effects, these stress release mechanisms may be activated slightly earlier/later in MLIP-MD (Fig.~\ref{FIG: validation}a: example of local stress drop at $\varepsilon=0.14$ and $\varepsilon=0.16$ during MLIP-MD and {\it{ab initio}} MD, respectively).
Evaluation of extrapolation grades (Fig.~\ref{FIG: validation}b: example of room-temperature tensile test for Ti$_2$AlB$_2$) indicates interpolation ($\gamma\leq1$) or accurate extrapolation ($\gamma<2$; in line with Refs.~\cite{lin2024machine,podryabinkin2023mlip,shapeev2020elinvar}), consistent with low statistical errors of MLIP-predicted energies, forces, and stresses (Fig.~\ref{FIG: validation}b: example of room-temperature tensile test for Ti$_2$AlB$_2$). 

Extensive validation of the the here-developed MLIPs (Fig.~\ref{FIG: validation}f--q) supports their transferability to mechanical---especially tensile---simulations for single-crystal MAB phases.
Importantly, no bias is shown with respect to certain phase, temperature, or loading condition (Fig.~\ref{FIG: validation}f--q).
Extrapolation grades underscore accurate extrapolation during all mechanical tests (Fig.~\ref{FIG: validation}f,j,n: $\gamma\leq1.5$) and the maximum stress error is $<2$~GPa (Fig.~\ref{FIG: validation}i,m,q).
Validation against compression tests (with up to 5~\% compression) indicates that our MLIPs can capture environments with small-to-medium compressive stresses, which may be relevant for Poisson's contraction during our targeted nanoscale tensile simulations.

\begin{landscape}
\onecolumn
\begin{adjustwidth}{-1.5in}{-1.85in} 
\begin{table*}[!htbp]
\centering
\fontsize{8}{10}\selectfont
\setlength{\tabcolsep}{2pt}
\caption{ \footnotesize
{\bf{Structural and elastic properties of the nine studied MABs predicted by MLIP(Ti-MABs), MLIP(Ta-MABs), and MLIP(W-MABs)}}.
The orthorhombic 222 ({\bf{M}}AlB), 212 ({\bf{M}}$_2$AB$_2$), and 314 ({\bf{M}}$_3$AB$_4$) type MABs, with A$=$Al, were modelled using $\approx{10^3}$-atom supercells (see the Methods).
MLIP-predicted lattice parameters ($a$, $b$, $c$) and per-atom volumes ($V$) are shown at $T\in\{0,300,1200\}$~K and their difference from similarly derived DFT/{\it{ab initio}} MD values is marked by $\Delta_{\text{ref-us}}$. 
Elastic constants, $C_{ij}$ (evaluated from second-order polynomial fit of the stress/strain data from uniaxial tensile or shear test, using 0--6\% strains with a 2\%-step), are shown together with the polycrystalline bulk ($B$), shear ($G$), Young's modulus ($E$), and Poisson's ratio ($\nu$).
For $C_{11}$, $C_{22}$, and $C_{33}$, differences from similarly calculated DFT/{\it{ab initio}} MD values is marked by $\Delta_{\text{ref-us}}$.   
} 
\label{TAB: aLat, Cij}
\begin{tabular}{ccccccccccccccccccccccc}
    \hline
    \hline
    \multicolumn{1}{c}{{\bf{M}}} & \multicolumn{1}{|c}{\bf MAB} & \multicolumn{1}{|c|}{\bf Temp.} & \multicolumn{4}{c|}{\bf Lattice parameters [\AA]} & \multicolumn{2}{c|}{\bf Volume [\AA$^3$/at.]} & \multicolumn{14}{c}{\bf Elastic constants and moduli [GPa / unitless: $\nu$] } \\ 
     &  \multicolumn{1}{|c}{\bf phase} & \multicolumn{1}{|c|}{$T$~{\bf{[K]}}} & $a$ & $b$ & $c$ & \multicolumn{1}{c|}{\cellcolor{blue!10} $\Delta_{\text{ref-us}}$~[\%] } & $V$ & \multicolumn{1}{c|}{\cellcolor{blue!10} $\Delta_{\text{ref-us}}$~[\%] } & $C_{11}$ & $C_{22}$ & $C_{33}$ & \multicolumn{1}{c|}{\cellcolor{blue!10} $\Delta_{\text{ref-us}}$~[\%]} & $C_{12}$ & $C_{23}$ & $C_{23}$ & $C_{44}$ & $C_{55}$ & $C_{66}$ & $B$ & $G$ & $E$ & $\nu$  \\
    \hline
    \hline
    {\bf{Ti}} & {\bf{222}}    & 0 &  3.283 & 3.049 & 14.648 & \cellcolor{blue!10}$\{0.0, 0.1, 0.2\}$ & 12.2  & \cellcolor{blue!10} $0.1$ & 315 & 245 & 245 & \cellcolor{blue!10}$\{1.4,4.9,9.8\}$ & 70 & 81 & 92 & 120 & 133 & 164 & 143 & 117 & 275 & 0.18 \\ 
                    & & & 3.297~\cite{liu2020new} & 3.049~\cite{liu2020new} & 14.641~\cite{liu2020new} & & 12.3~\cite{liu2020new} & & 352~\cite{liu2020new} & 273~\cite{liu2020new} & 202~\cite{liu2020new} & & 74~\cite{liu2020new} & 83~\cite{liu2020new} & 92~\cite{liu2020new} & 122~\cite{liu2020new} & 130~\cite{liu2020new} & 158~\cite{liu2020new} & 145~\cite{liu2020new} & 116~\cite{liu2020new} & 274~\cite{liu2020new} & 0.18~\cite{liu2020new} \\ 
                    & & & 3.281~\cite{koutna2024phase} & 3.056~\cite{koutna2024phase} & 14.691~\cite{koutna2024phase} & & 12.3~\cite{koutna2024phase} & & 350~\cite{koutna2024phase} & 263~\cite{koutna2024phase} & 191~\cite{koutna2024phase} & & 76~\cite{koutna2024phase} & 81~\cite{koutna2024phase} & 99~\cite{koutna2024phase} & 119~\cite{koutna2024phase} & 130~\cite{koutna2024phase} & 139~\cite{koutna2024phase} & 143~\cite{koutna2024phase} & 109~\cite{koutna2024phase} & 261~\cite{koutna2024phase} & 0.20~\cite{koutna2024phase} \\ 
                    & & 300 & 3.296 & 3.058 & 14.687 & \cellcolor{blue!10} $\{1.0,0.2, 0.8 \}$ & 12.3 & \cellcolor{blue!10} 0.0 & 304 & 235 & 190 & \cellcolor{blue!10} $\{2.5,3.1,12.8\}$ & 71 & 82 & 102 & 153 & 129 & 113 & 136 & 104 & 248 & 0.19 \\ 
                    & & 1200 & 3.329 & 3.090 & 14.880 & \cellcolor{blue!10} $\{0.0 ,0.0 ,0.2 \}$ & 12.8 & \cellcolor{blue!10} $0.2$ & 273 & 199 & 127 & \cellcolor{blue!10}$\{2.7,1.8,5.3\}$ & 70 & 73 & 82 & 125 & 102 & 93 & 112 & 81 & 195 & 0.21 \\ 
    \hline
    & {\bf{212}} & 0 & 3.050 & 3.295 & 11.298 & \cellcolor{blue!10} $\{0.0,0.2,0.0 \}$ & 11.4 & \cellcolor{blue!10} 0.2 & 470 & 340 & 305 & \cellcolor{blue!10}$\{3.6,1.3,11.7\}$ & 98 & 86 & 127 & 209 & 196 & 223 & 241 & 150 & 371 & 0.18 \\ 
                            & & &  3.046~\cite{liu2020new} & 3.308~\cite{liu2020new} & 11.320~\cite{liu2020new} & & 11.4~\cite{liu2020new} & & 406~\cite{liu2020new} & 307~\cite{liu2020new} & 270~\cite{liu2020new} & & 91~\cite{liu2020new} & 	80~\cite{liu2020new} & 108~\cite{liu2020new} & 176~\cite{liu2020new} & 165~\cite{liu2020new} & 204~\cite{liu2020new} & 170~\cite{liu2020new} & 150~\cite{liu2020new} & ~347\cite{liu2020new} & 0.16~\cite{liu2020new} \\ 
                            & & & 3.049~\cite{koutna2024phase} & 3.309~\cite{koutna2024phase} & 11.327~\cite{koutna2024phase} & & 12.4~\cite{koutna2024phase} & & 403~\cite{koutna2024phase} & 308~\cite{koutna2024phase} & 266~\cite{koutna2024phase} & & 92~\cite{koutna2024phase} & 80~\cite{koutna2024phase} & 107~\cite{koutna2024phase} & 171~\cite{koutna2024phase} & 164~\cite{koutna2024phase} & 204~\cite{koutna2024phase} & 169~\cite{koutna2024phase} & 148~\cite{koutna2024phase} & 344~\cite{koutna2024phase} & 0.16~\cite{koutna2024phase} \\ 
                           & & 300 & 3.057 & 3.306 & 11.342 & \cellcolor{blue!10} $\{0.0 ,0.1 ,0.1 \}$ & 11.5 & \cellcolor{blue!10} 0.0  & 375 & 296 & 238 & \cellcolor{blue!10}$\{3.5,1.5,3.3\}$ & 88 & 81 & 118 & 166 & 161 & 185 & 163 & 136 & 320 & 0.17 \\ 
                           & & 1200 & 3.084 & 3.335 & 11.466 & \cellcolor{blue!10} $\{-0.1 ,0.1 ,0.0 \}$ &  11.8 & \cellcolor{blue!10} $0.0$ & 328 & 259 & 189 & \cellcolor{blue!10}$\{3.9,4.3,0.3\}$ & 88 & 80 & 100 & 139 & 137 & 156 & 143 & 113 & 268 & 0.19 \\ 
    \hline
     & {\bf{314}}   & 0 & 3.039 & 3.274 & 8.256 & \cellcolor{blue!10} $\{0.0 ,0.2 ,0.1 \}$ & 10.3 & \cellcolor{blue!10} 0.1 & 470 & 340 & 305 & \cellcolor{blue!10}$\{17.6,13.0,2.4\}$ & 98 & 86 & 127 & 209 & 340 & 305 & 191 & 171 & 396 & 0.16\\ 
                            & & 300 & 3.044 & 3.287 & 8.282 & \cellcolor{blue!10} $\{0.0 ,0.1 , 0.2 \}$ & 10.4 & \cellcolor{blue!10} $0.1$ & 448 & 332 & 286 & \cellcolor{blue!10}$\{2.6,3.7,9.4\}$ & 90 & 81 & 116 & 198 & 186 & 213 & 181 & 165 & 379 & 0.15 \\ 
                            & & 1200 & 3.066 & 3.317 & 8.362 & \cellcolor{blue!10} $\{0.1 ,0.1 ,-0.1 \}$ & 10.6 & \cellcolor{blue!10} $0.1$ & 410 & 315 & 244 & \cellcolor{blue!10}$\{3.7,7.8,7.2\}$ & 93 & 85 & 117 & 172 & 160 & 198 & 171 & 142 & 334 & 0.17\\ 
    \hline
    \hline
    {\bf{Ta}} & {\bf{222}}  & 0 & 3.328 & 3.106 & 14.592 &  \cellcolor{blue!10} $\{0.5 ,0.0 ,0.3 \}$ & 12.6 & \cellcolor{blue!10} 0.1 & 331 & 288 & 272 & \cellcolor{blue!10}$\{15.4,13.0,2.4\}$ & 128 & 169 & 154 & 173 & 174 & 164 & 198 & 120 & 300 & 0.25 \\ 
                    & & & 3.351~\cite{liu2020new} & 3.107~\cite{liu2020new} & 14.576~\cite{liu2020new} & & 12.6~\cite{liu2020new} & & 369~\cite{liu2020new} & 321~\cite{liu2020new} & 278~\cite{liu2020new} & & 131~\cite{liu2020new} & 	116~\cite{liu2020new} & 138~\cite{liu2020new} & 187~\cite{liu2020new} & 163~\cite{liu2020new} & 181~\cite{liu2020new} & 192~\cite{liu2020new} & 	138~\cite{liu2020new} & 333~\cite{liu2020new} & 0.21~\cite{liu2020new} \\
                    & & & 3.336~\cite{koutna2024phase} & 3.107~\cite{koutna2024phase} & 14.662~\cite{koutna2024phase} & & 12.7~\cite{koutna2024phase} & & 354~\cite{koutna2024phase} & 289~\cite{koutna2024phase} & 271~\cite{koutna2024phase} & & 140~\cite{koutna2024phase} & 119~\cite{koutna2024phase} & 183~\cite{koutna2024phase} & 183~\cite{koutna2024phase} & 163~\cite{koutna2024phase} & 181~\cite{koutna2024phase} & 193~\cite{koutna2024phase} & 128~\cite{koutna2024phase} & 315~\cite{koutna2024phase} & 0.23~\cite{koutna2024phase} \\ 
                    & & 300 & 3.334 & 3.121 & 14.653 &  \cellcolor{blue!10} $\{0.3 ,0.3,0.1 \}$ & 12.7 & \cellcolor{blue!10} 0.1 & 306 & 273 & 260 & \cellcolor{blue!10}$\{11.2,10.1,3.4 \}$ & 133 & 150 & 112 & 162 & 162 & 150 & 180 & 116 & 186 & 0.24 \\
                    & & 1200 & 3.354 & 3.151 & 14.786 & \cellcolor{blue!10} $\{0.2 ,0.3 ,0.0 \}$ & 13.0 & \cellcolor{blue!10} $0.0$ & 254 & 247 & 217 & \cellcolor{blue!10}$\{8.6,11.6,5.0\}$ & 116 & 135 & 100 & 136 & 131 & 125 & 156 & 96 & 239 & 0.25 \\ 
    \hline
    & {\bf{212}}   & 0 & 3.098 & 3.340 & 11.615 & \cellcolor{blue!10} $\{0.7,0.1,0.9\}$ & 12.0 & \cellcolor{blue!10} 0.4 & 312 & 287 & 223 & \cellcolor{blue!10}$\{2.3,2.5,5.8\}$ & 109 & 131 & 173 & 134 & 111 & 176 & 182 & 98 & 248 & 0.27 \\ 
                            & & 300 & 3.107 & 3.344 & 11.678 & \cellcolor{blue!10} $\{0.3,0.3,0.1\}$ & 12.1 & \cellcolor{blue!10}  $0.5$ & 315 & 255 & 203 & \cellcolor{blue!10}$\{9.1,2.2,1.5\}$ & 123 & 139 & 134 & 123 & 104 & 166 & 171 & 93 & 238 & 0.27 \\ 
                            & & 1200 & 3.137 & 3.352 & 11.858 & \cellcolor{blue!10} $\{0.2,0.1,0.2\}$ & 12.5 & \cellcolor{blue!10} $0.0$ & 290 & 248 & 169 & \cellcolor{blue!10}$\{2.3,0.8,14.4\}$ & 147 & 148 & 131 & 106 & 92 & 143 & 166 & 76 & 198 & 0.30\\ 
    \hline
    & {\bf{314}}  & 0 &  3.101 & 3.308 & 8.502 & \cellcolor{blue!10} $\{0.2,0.2 ,0.2 \}$ & 10.9 & \cellcolor{blue!10}  0.2 & 482 & 341 & 285 & \cellcolor{blue!10}$\{0.5,14.4,11.7\}$ & 180 & 187 & 149 & 165 & 139 & 215 & 230 & 135 & 338 & 0.25 \\ %
                          &  & 300 & 3.110 & 3.316 & 8.529 & \cellcolor{blue!10} $\{0.1,0.3 ,0.4 \}$ & 11.0 & \cellcolor{blue!10} 0.2 & 449 & 326 & 267 & \cellcolor{blue!10}$\{2.7,11.4,4.1\}$ & 161 & 162 & 132 & 160 & 132 & 207 & 211 & 131 & 326 & 0.24 \\
                          &  & 1200 & 3.141 & 3.331 & 8.616 & \cellcolor{blue!10} $\{0.2 ,0.1 ,0.2 \}$ & 11.3 & \cellcolor{blue!10} $0.1$ & 382 & 324 & 228  & \cellcolor{blue!10}$\{6.6,3.6,11.2\}$ & 155 & 143 & 141 & 138 & 117 & 182 & 196 & 113 & 284 & 0.26\\
    \hline
    \hline
    {\bf{W}} & {\bf{222}}  & 0 & 3.207 & 3.121 & 13.943 & \cellcolor{blue!10} $\{0.4,0.2,0.1\}$ & 11.6 & \cellcolor{blue!10} 0.1 & 378 & 333 & 389 & \cellcolor{blue!10}$\{ 15.5,6.2,11.4\}$ & 203 & 133 & 144 & 194 & 206 & 159 & 229 & 144 & 356 & 0.24 \\ 
                    & & & 3.217~\cite{liu2020new} & 3.118~\cite{liu2020new} & 13.985~\cite{liu2020new} & & 11.7~\cite{liu2020new}& & 405~\cite{liu2020new} & 365~\cite{liu2020new} & 349~\cite{liu2020new} & & 189~\cite{liu2020new} &  143~\cite{liu2020new} & 160~\cite{liu2020new} & 183~\cite{liu2020new} & 195~\cite{liu2020new} & 171~\cite{liu2020new} & 233~\cite{liu2020new} & 145~\cite{liu2020new} & 361~\cite{liu2020new} & 0.24~\cite{liu2020new} \\   
                    & & & 3.223~\cite{koutna2024phase} & 3.122~\cite{koutna2024phase} & 13.990~\cite{koutna2024phase} & & 11.7~\cite{koutna2024phase} & & 396~\cite{koutna2024phase} & 385~\cite{koutna2024phase} & 356~\cite{koutna2024phase} & & 183~\cite{koutna2024phase} & 142~\cite{koutna2024phase} & 155~\cite{koutna2024phase} & 181~\cite{koutna2024phase} & 196~\cite{koutna2024phase} & 170~\cite{koutna2024phase} & 229~\cite{koutna2024phase} & 146~\cite{koutna2024phase} & 361~\cite{koutna2024phase} & 0.24~\cite{koutna2024phase} \\ 
                    & & 300 & 3.218 & 3.128 & 13.953 & \cellcolor{blue!10} $\{0.2,0.4,0.7 \}$ & 11.7 & \cellcolor{blue!10}  0.0 & 357 & 339 & 342 & \cellcolor{blue!10}$\{9.4,5.2,6.8\}$ & 196 & 118 & 143 & 188 & 153 & 188 & 215 & 136 & 337 & 0.24 \\
                                     & & & 3.207~\cite{roy2023low} & 3.104~\cite{roy2023low} & 13.924~\cite{roy2023low} & & 11.6~\cite{roy2023low}  \\  
                    & & 1200 & 3.238 & 3.147 & 14.147 & \cellcolor{blue!10} $\{0.3,0.1,0.1 \}$ & 12.0  & \cellcolor{blue!10}  0.1 & 334 & 329 & 334 & \cellcolor{blue!10}$\{7.6,1.0,11.1\}$ & 195 & 129 & 154 & 148 & 160 & 128 & 203 & 109 & 277 & 0.25  \\ 
    \hline
    & {\bf{212}} & 0~K & 3.081 & 3.140 & 11.505 & \cellcolor{blue!10} $\{0.1,0.0,0.2\}$  & 11.1 & \cellcolor{blue!10}  0.2 & 556 & 494 & 370 & \cellcolor{blue!10}$\{13.3,6.6,1.5\}$ & 196 & 158 & 194 & 176 & 170 & 234 & 275 & 169 & 422 & 0.24\\ 
                        & & & 3.088~\cite{liu2020new} & 3.139~\cite{liu2020new} & 	11.566~\cite{liu2020new} & & 11.2~\cite{liu2020new}& & 511~\cite{liu2020new} & 474~\cite{liu2020new} & 365~\cite{liu2020new} & & 187~\cite{liu2020new} & 154~\cite{liu2020new} & 179~\cite{liu2020new} & 147~\cite{liu2020new} & 115~\cite{liu2020new} & 163~\cite{liu2020new} & 262~\cite{liu2020new} & 139~\cite{liu2020new} & 354~\cite{liu2020new} & 0.28~\cite{liu2020new} \\ 
                      &  & 300 & 3.089 & 3.149 & 11.545 & \cellcolor{blue!10} $\{0.2,0.0,0.1 \}$ & 11.2 & \cellcolor{blue!10} 0.1 & 487 & 458 & 334 & \cellcolor{blue!10}$\{4.4,2.2,3.4\}$ & 175 & 148 &183 & 166 & 162 & 221 & 251 & 156 & 388 & 0.24\\
                      &  & 1200 & 3.104 & 3.173 & 11.666 & \cellcolor{blue!10} $\{0.1,0.1,0.0 \}$ & 11.5 & \cellcolor{blue!10} $0.0$& 398 & 374 & 264 & \cellcolor{blue!10}$\{2.7,3.2,9.1\}$ & 166 & 147 & 148 & 147 & 142 & 193 & 213 & 127 & 319 & 0.25 \\ 
    \hline
    & {\bf{314}} & 0 & 3.079 & 3.161 & 8.390 & \cellcolor{blue!10} $\{0.5,0.6,0.0\}$ &  10.2 & \cellcolor{blue!10} 1.1 & 498 & 322 & 342 & \cellcolor{blue!10}$\{5.8,14.8,10.4\}$ & 174 & 142 & 149 & 170 & 167 & 200 & 228 & 148 & 364 & 0.23 \\ 
                      &  & 300 & 3.085 & 3.174 & 8.418 &  \cellcolor{blue!10} $\{0.3, 0.2, 0.3\}$ & 10.3 & \cellcolor{blue!10} 0.2 & 490 & 377 & 361 & \cellcolor{blue!10}$\{1.0,13.8,10.8 \}$ & 212 & 185 & 192 & 155 & 163 & 180 & 265 & 138 & 351 & 0.27\\ 
                      &  & 1200 & 3.112 & 3.204 & 8.495 & \cellcolor{blue!10}$\{0.1, 0.1, 0.1\}$ & 10.6 &\cellcolor{blue!10} $0.2$ & 394 & 299 & 285 & \cellcolor{blue!10}$\{6.7,14.5,3.9\}$ & 174 & 142 & 119 & 132 & 143 & 160 & 201 & 119 & 298 & 0.25 \\ 
    \hline
    \hline
\end{tabular}
\end{table*}
\end{adjustwidth}
\end{landscape}


\twocolumn

\subsection{Mechanical response of single-crystal MAB phases at the nanoscale}
\begin{figure*}[h!t!]
    \centering
    \includegraphics[width=2\columnwidth]{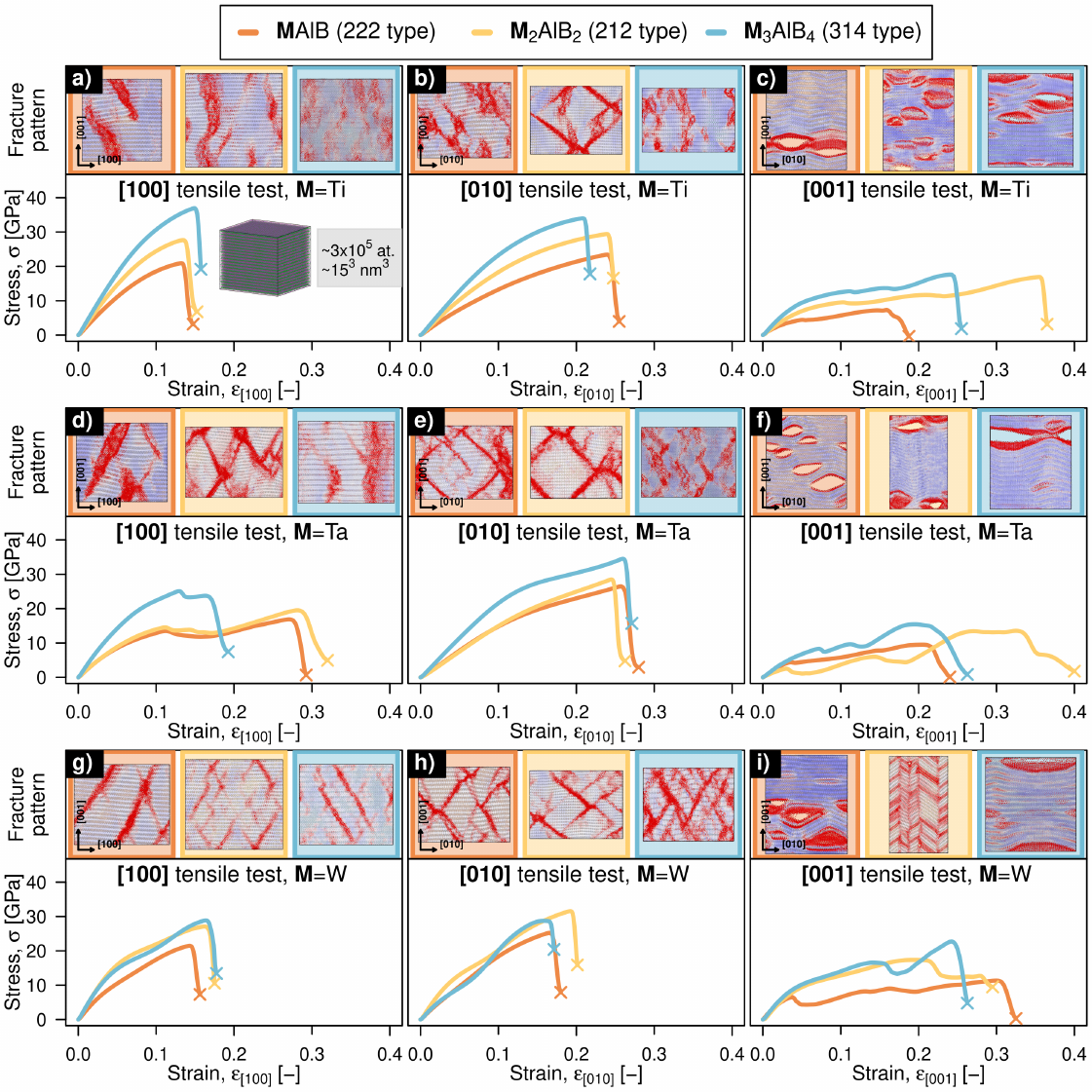}
    \caption{
    \small
    {\bf{Stress/strain curves for the nine studied MABs, M$=(\text{Ti, Ta, W})$, A$=$Al, subject to uniaxial tensile deformation simulated by nanoscale MLIP-MD at 300~K}}. 
    During each mechanical test, tensile strain ($\varepsilon$) was applied to initially defect-free supercells along the [100], [010], or [001] direction.
    The stress ($\sigma$) represents stress tensor component in the loaded direction (stresses normal to this direction are nullified due to Poisson's effect).
    Panels (a--c), (d--f), and (g--i) depict results for Ti-, Ta-, and W-based MABs.  
    The orange, yellow, and blue line marks the 222, 212, and 314 type phase, respectively, each modelled by $\approx{300,000}$-atom supercell ($\approx{15^3}$~nm$^3$).
    Snapshots on the top of each panel illustrate fracture patterns, visualised by volumetric strain distribution (blue$\to$white$\to$red; with red denoting highly tensile-strained regions), where the frame colour marks the phase prototype. 
    Note that determining exact fracture strains may be ambiguous and we only consider stress/strain data corresponding to reliable extrapolation grades ($\gamma<10$).
    }
\label{FIG: stress/strain}
\end{figure*}

Validation of the here-developed MLIPs indicates that they can tackle various strain conditions far from elastic regime, particularly those relevant for tensile deformation.
Employing these MLIPs, we simulate tensile tests for single-crystal MABs at length scales allowing for strain-induced nucleation of extended defects (Fig.~\ref{FIG: stress/strain}). 
Specifically, the nine MABs previously modelled in $\approx{10}^3$-atom supercells are now represented by supercells with $\approx{300,000}$ atoms ($\approx{15^3}$~nm$^3$).
Although such simulations cannot be reproduced by {\it{ab initio}} MD, their reliability is indicated by (i) the MLIP's agreement with relevant atomic-scale {\it{ab initio}} MD calculations (Tab.~\ref{TAB: aLat, Cij}, Fig.~\ref{FIG: validation}), (ii) negligible changes of size-independent material's properties (e.g., structural and elastic data from Tab.~\ref{TAB: aLat, Cij} remaining unchanged for supercells with 10$^1$--10$^6$ atoms), (iii) low extrapolation grades.
The last mentioned can be evaluated also for nanoscale supercells under far-from-equilibrium conditions and indicate reliable extrapolation ($\gamma<10$) of mechanical tests presented below.

Stress/strain curves derived from room-temperature tensile tests for the nine studied MABs (Fig.~\ref{FIG: stress/strain}) show qualitatively different response to tensile loads parallel to M--B/A layers, i.e., along $[100]\parallel a$ and $[010]\parallel b$, and orthogonal to them, i.e., along $[001]\parallel c$.  
While the former is relatively stiff, the latter is initially softer and then  exhibits strain hardening, hence resembling behaviour typical for metals. 
Strongly-directional mechanical response of MAB phases, suggested already by elastic constants calculations (Tab.~\ref{TAB: aLat, Cij}, Refs.~\cite{koutna2024phase,liu2020new,dai2018first}), therefore extends also beyond the linear-elastic regime. 
Qualitative differences between in-plane and out-of-plane mechanical behaviour are further underpinned by deformation mechanisms (see snapshots at fracture in Fig.~\ref{FIG: stress/strain}). 
Specifically, in-plane tensile response is characterised by layer slipping, interpenetration, and crack nucleation diagonally across the layers or along the [001] direction.
Out-of-plane tensile loading causes layer bending and formation of kink bands followed by delamination.

Starting with {\bf{Ti-based MABs}} (Fig.~\ref{FIG: stress/strain}a--c, Tab.~\ref{TAB: mechanical properties}),
three hypotheses can be drawn.
{\bf{(i)}} In-plane tensile strengths, $\sigma_{\text{[100]}}$ and $\sigma_{\text{[010]}}$, of any given phase prototype are very close and notably above $\sigma_{\text{[001]}}$; 
e.g., $\sigma_{\text{[100]}}\approx{28.0}$~GPa of Ti$_2$AlB$_2$ is similar to $\sigma_{\text{[010]}}\approx{29.8}$~GPa but significantly above $\sigma_{\text{[001]}}\approx{17.2}$~GPa.
{\bf{(ii)}} The strain at fracture, however, is higher subject to [010] loading, rendering the highest toughness, $U_{\text{[010]}}$, in the [010] direction; e.g., $U_{\text{[010]}}\approx{4.60}$~GPa of Ti$_2$AlB$_2$ surpasses its  $U_{\text{[100]}}\approx{2.34}$~GPa.
{\bf{(ii)}} Both ideal tensile strength and toughness increase for TiAlB$\to$Ti$_2$AlB$_2$$\to$Ti$_3$AlB$_4$, irrespective of the loading direction.
This may seem intuitive for strength---mirroring the increased fraction of the harder ceramic material---but less for toughness.  
Examining structural changes at various deformation stages, we observe earlier layer bending, delamination, or layer interpenetration for TiAlB and Ti$_2$AlB$_2$, in contrast with Ti$_3$AlB$_4$, and  their lower toughness may therefore originate from premature plasticity.

\begin{table*}[!htbp]
\centering
\footnotesize
\setlength{\tabcolsep}{4pt}
\caption{ \footnotesize
{\bf{Mechanical properties of the nine studied MABs, M$=(\text{Ti, Ta, W})$, A$=$Al, derived from nanoscale MLIP-MD simulations (Fig.~\ref{FIG: stress/strain})}}.
Theoretical tensile strengths---$\sigma_{\text{[100]}}^{\text{max}}$, $\sigma_{\text{[010]}}^{\text{max}}$, and $\sigma_{\text{[001]}}^{\text{max}}$---were evaluated as the maximum stress reached during the [100], [010], and [001] tensile test, respectively, where their mean value is denoted by $\overline{\sigma}_{\text{max}}$ (shown together with standard deviation indicating the extent of anisotropy). 
Theoretical tensile toughness values---$U_{\text{[100]}}^{\text{max}}$, $U_{\text{[010]}}^{\text{max}}$, and $U_{\text{[001]}}^{\text{max}}$---were evaluated by integrating the corresponding stress/strain curve until the maximum stress point.
The mean value is denoted by $\overline{U}$ (shown together with standard deviation).
As the supercells were initially defect-free, all values represent ideal upper bounds of strength and toughness attainable in single-crystal MABs with initially no defects.
} 
\label{TAB: mechanical properties}
\begin{tabular}{cccccccccccc}
    \hline
    \hline
    \multicolumn{1}{c}{{\bf{M}}} & \multicolumn{1}{|c}{\bf MAB} & \multicolumn{1}{|c}{\bf Number of} & \multicolumn{1}{|c|}{\bf Temp.~{\bf{[K]}}} & \multicolumn{4}{c|}{\bf Theoretical tensile strength [GPa]} & \multicolumn{4}{c}{\bf  Theoretical tensile toughness [GPa]} \\ 
     &  \multicolumn{1}{|c}{\bf phase} & \multicolumn{1}{|c}{\bf atoms } & \multicolumn{1}{|c|}{$T$} & $\sigma_{\text{[100]}}^{\text{max}}$ & $\sigma_{\text{[010]}}^{\text{max}}$ & $\sigma_{\text{[010]}}^{\text{max}}$ & \multicolumn{1}{c|}{ $\overline{\sigma}^{\text{max}}$ } & $U_{\text{[100]}}$ & $U_{\text{[010]}}$ & $U_{\text{[001]}}$ & $\overline{U}$ \\
    \hline
    {\bf{Ti}} & {\bf{222}} & 243,000 & 300 & 21.3 & 23.6 & 7.7 & 17.5$\pm$8.6 & 1.79 & 3.42 & 0.82 & 2.00$\pm$1.30  \\ 
    & {\bf{212}} & 286,650 & 300 & 28.0 & 29.8 & 17.2 & 25.0$\pm$6.8 & 2.34 & 4.60 & 3.86 &  3.60$\pm$1.20 \\ 
     & {\bf{314}} & 294,914 & 300 & 37.3 & 34.4 & 17.8 & 29.9$\pm$10.5 & 3.47 & 4.59 & 2.92 &  3.70$\pm$0.90  \\ 
    \hline
    {\bf{Ta}} & {\bf{222}}  & 243,000 & 300 & 17.2 & 26.7 & 9.6 & 17.8$\pm$8.5 & 3.19 & 4.16 & 1.26 &  2.90$\pm$1.50 \\ 
    & {\bf{212}} & 286,650 & 300 & 19.7 & 28.8 & 13.7 & 20.7$\pm$7.6 & 3.64 & 4.06 & 2.18 & 3.30$\pm$1.00 \\ 
    & {\bf{314}} & 294,914 & 300 & 25.5 & 34.9& 15.6 & 25.3$\pm$9.7 & 2.85 & 5.86 & 1.76 & 3.49$\pm$2.12 \\ 
    \hline
    {\bf{W}} & {\bf{222}} & 243,000 & 300 & 21.9 & 25.5 & 11.6 & 19.7$\pm$7.2 & 1.91 & 2.50 & 2.37 & 2.30$\pm$0.30 \\ 
    & {\bf{212}} & 286,650 & 300 & 27.4 & 32.1 & 17.4 & 25.6$\pm$7.5 & 3.01 & 3.80 & 2.21 & 3.00$\pm$0.80 \\ 
    & {\bf{314}} & 294,914 & 300 & 29.3 & 29.1 & 23.0 & 27.1$\pm$3.6 & 3.02 & 2.51 & 3.31 & 2.90$\pm$0.40 \\ 
    \hline
\end{tabular}
\end{table*}

For {\bf{Ta-based MABs}} (Fig.~\ref{FIG: stress/strain}d--f, Tab.~\ref{TAB: mechanical properties}), MLIP-MD results render the following findings.
{\bf{(i)}} The [010] direction is the hardest and the toughest:  $\sigma_{\text{[100]}}>\sigma_{\text{[010]}}>\sigma_{\text{[001]}}$ (at variance with fairly basal-plane-isotropic Ti-MABs) and analogically for toughness. 
E.g., $\sigma_{\text{[100]}}\approx{29.7}$~GPa of Ta$_2$AlB$_2$ exceeds its  $\sigma_{\text{[010]}}\approx{19.7}$~GPa as well as $\sigma_{\text{[001]}}\approx{13.7}$~GPa.
{\bf{(ii)}} Not only the [001] but also the [100] direction exhibits strain hardening, contrarily to Ti-MABs.
Also stress/strain data for [010] deformation indicate strain hardening, though less pronounced compared with the [100] and [010] loading conditions. 
Similarly to Ti-MABs, {\bf{(iii)}} ideal tensile strength increases upon increasing the fraction of the ceramic-like component, i.e., for TaAlB$\to$Ta$_2$AlB$_2$$\to$Ta$_3$AlB$_4$. 
The trend for toughness is almost like this, with the exception of [100] and [001] deformation during which Ta$_2$AlB$_2$ provides the highest toughness. 

{\bf{W-based MABs}} (Fig.~\ref{FIG: stress/strain}g--i, Tab.~\ref{TAB: mechanical properties}) show fairly similar response to in-plane, [100] and [010], tensile strains, contrarily to Ta-MABs, but also Ti-MABs.
Strain hardening is predicted subject to [001] tensile loading, comparable to $\text{M}=\{$Ti, Ta$\}$.
Noteworthy is the nucleation of twin boundaries in W$_2$AlB$_2$ (isoelectronic to the experimentally known Cr$_2$AlB$_2$~\cite{berastegui2020magnetron,kota2018magnetic,ade2015ternary}) during [001] tensile test, indicating superior resistance to crack initiation and propagation.
The experimentally known WAlB does not particularly stand out, showing rather low strength ($\sigma\approx{11.6}$--25.5~GPa) and moderate toughness ($U\approx{1.91}$--2.50~GPa).

Among the studied MABs, Ti$_3$AlB$_4$ exhibits the highest theoretical tensile strength (Tab.~\ref{TAB: mechanical properties}: $\sigma_{\text{[100]}}\approx{37.3}$~GPa; $\sigma_{\text{[010]}}\approx{34.4}$~GPa), which is about 45~\% below [$\overline{1}2\overline{1}0$] tensile strength of a typical hard ceramic, TiB$_2$~\cite{lin2024machine}.
While Ti$_3$AlB$_4$ has not been synthesised, it has the lowest formation energy, $E_f$, among various mechanically and dynamically stable Ti-MABs (namely, $E_f$ of Ti$_2$AlB$_2$ and TiAlB is 0.07 and 0.24~eV/at. higher, respectively)~\cite{koutna2024phase}.
Other MABs with high strength include Ti$_2$AlB$_2$, Ta$_3$AlB$_4$, and W$_3$AlB$_4$.

The overall highest toughness is exhibited by Ta$_3$AlB$_4$ (Tab.~\ref{TAB: mechanical properties}: $U_{\text{[010]}}\approx{5.86}$~GPa), which is about 55~\% above that of TiB$_2$~\cite{lin2024machine}.
Ta$_3$AlB$_4$ is energetically close to the ground-state Ta$_2$AlB$_2$ (0.01~eV/at. higher in $E_f$)~\cite{koutna2024phase}.  
Note, however, that our toughness ($U$) expresses the energy consumed by the material until the global stress maximum, not until the fracture point which may be ambiguous. 
Consequently, also W$_2$AlB$_2$ could possess superior toughness, as hinted by the structure snapshot at the onset of fracture (Fig.~\ref{FIG: stress/strain}i) showing nucleation of V-shaped defects, later identified as twin boundaries.

\subsubsection{Response to in-plane tension: shear bands}
\begin{figure*}[h!t!]
    \centering
    \includegraphics[width=1.75\columnwidth]{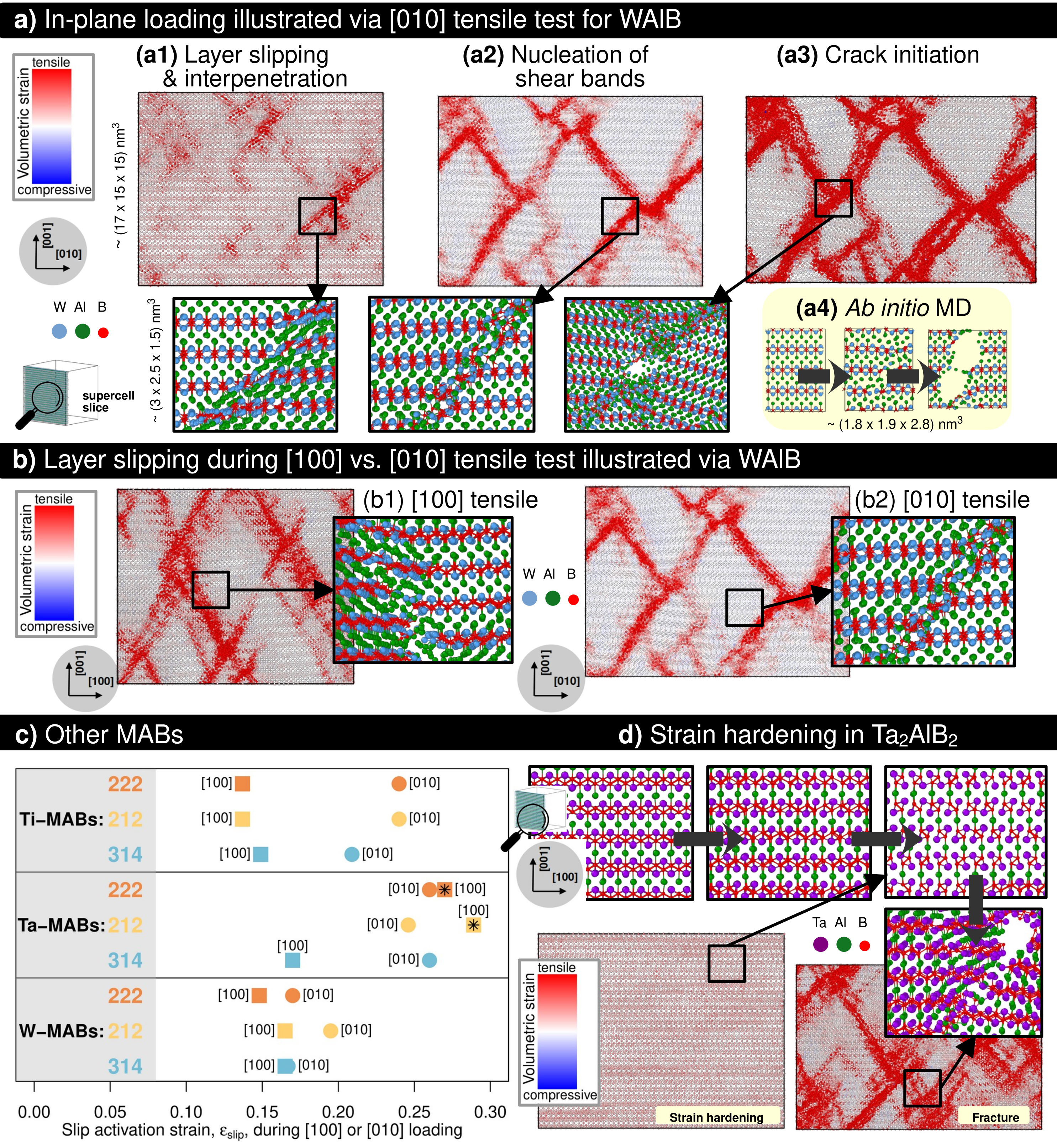}
    \caption{
    \small
    {\bf{Deformation mechanisms of MAB phases during in-plane loading, derived form MLIP-MD room-temperature uniaxial tensile tests}}.
    (a) Typical strain-induced structural changes illustrated by WAlB (stress/strain data in Fig.~\ref{FIG: stress/strain}h).
    The blue$\to$white$\to$red colouring mirrors distribution of the volumetric strain (compressive$\to$zero$\to$tensile) and highlights crack initiation patterns (red). 
    The magnifying glass icon marks snapshots that are local magnifications of the supercell, typically $\approx{(3\times3\times2)}$~nm$^3$. 
    (b) Layer slipping during [100] and [010] tensile tests illustrated by WAlB (stress/strain data in Fig.~\ref{FIG: stress/strain}g--h).
    (c) Slip activation strain during [100] and [010] tensile tests (stress/strain data in Fig.~\ref{FIG: stress/strain}), where the star for TaAlB and Ta$_2$AlB$_2$ indicates that these MABs exhibit strain hardening.
    (d) Illustration of strain hardening mechanism in Ta$_2$AlB$_2$ during [100] tensile test (stress/strain data in Fig.~\ref{FIG: stress/strain}d).
      }
\label{FIG: in-plane loading}
\end{figure*}

Tensile loading parallel to basal planes induces layer slipping, typically along the $\{110\}\langle001\rangle$-type slip system, followed by nucleation of shear bands, most often diagonally across M--B/A layers.
These phenomena---observed in all MABs---are exemplified by WAlB during room-temperature [010] tensile test (Fig.~\ref{FIG: in-plane loading}a).
The high strain concentration localized along shear bands (blue$\to$white$\to$red colour code in Fig.~\ref{FIG: in-plane loading}a) causes nucleation of X-shaped defects and void opening inside the supercell (see local magnification in Fig.~\ref{FIG: in-plane loading}a3).

\begin{figure*}[h!t!]
    \centering
    \includegraphics[width=1.75\columnwidth]{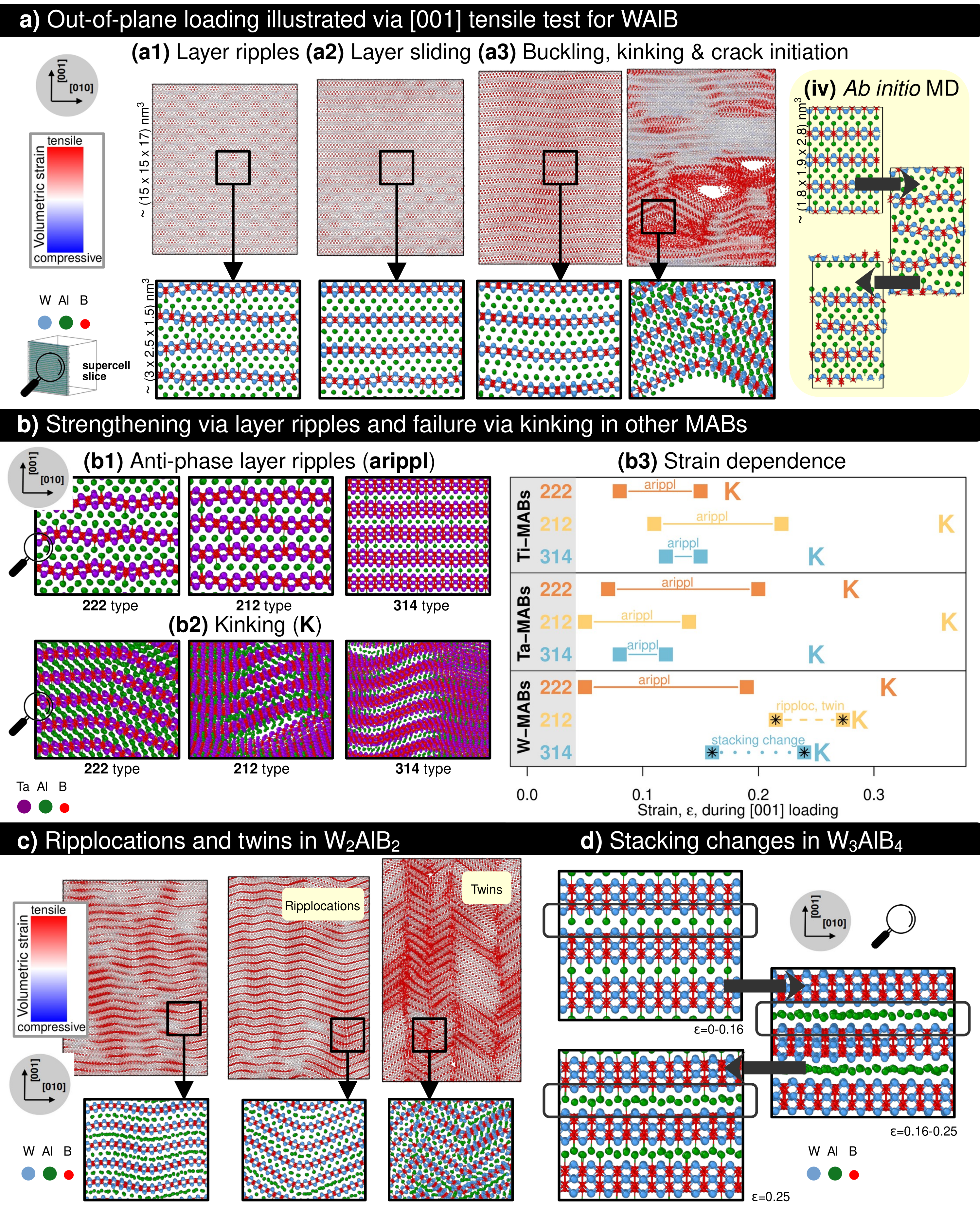}
    \caption{
    \small
    {\bf{Deformation mechanisms of MAB phases during loading normal to basal planes, derived form MLIP-MD room-temperature uniaxial tensile tests}}.
    (a) Typical strain-induced structural changes illustrated by WAlB (stress/strain data in Fig.~\ref{FIG: stress/strain}i). 
    The blue$\to$white$\to$red colouring mirrors distribution of the volumetric strain (compressive$\to$zero$\to$tensile) and highlights crack initiation patterns (red). 
    The magnifying glass icon marks snapshots that are local magnifications of the supercell, typically $\approx{(3\times3\times2)}$~nm$^3$.
    (b) Magnifications of layer ripples and kinks together with the corresponding strains for all MABs.
    (c) Snapshots of ripplocations and twins in W$_2$AlB$_2$.
    (d) Stacking faults in W$_4$AlB$_4$.
    }
\label{FIG: out-of-plane loading}
\end{figure*}

Mechanical failure by shear banding resembles behaviour of metals (e.g., Cu/Nb nanocomposites~\cite{chen2020effects}).
Despite no configurations with shear bands were explicitly trained on, already atomic-scale {\it{ab initio}} MD tensile tests used to validate our MLIPs showed M--B/A layer slipping and intermixing (Fig.~\ref{FIG: in-plane loading}a4), thus supporting nanoscale predictions.
Layer slipping and interpenetration is qualitatively similar for all studied MABs (see local magnifications in Fig.~\ref{FIG: in-plane loading}b).
The necessary activation strain, however, typically decreases during [100] tensile test compared with [010] tensile test, suggesting lower energetic costs of (110)[100] slip (Fig.~\ref{FIG: in-plane loading}c).
This effect is very pronounced for Ti-based MABs and gets less striking for their W-containing counterparts. 

TaAlB and Ta$_2$AlB$_2$ are exceptions: postponing the onset of (011)[100] slip during [100] tensile test beyond the activation strain of (101)[010] slip during [010] loading. 
This is achieved via small bonding rearrangements (particularly tilting) within the ceramic-like M--B layers (Fig.~\ref{FIG: in-plane loading}d), mirrored by work strengthening (stress/strain data in Fig.~\ref{FIG: stress/strain}d).
Contrarily to four and six valence electrons of Ti and W, the five valence electrons of Ta may be ``just right'' to activate these tiny rearrangements within the M--B network, hindering the onset of shear bands and fracture.
In-depth understanding, however, would require careful bonding analysis.

\subsubsection{Response to out-of-plane tension:\\ Layer buckling, kink bands, and twins}
Work hardening and high fracture strains shown in stress/strain curves for [001] tensile tests (Fig.~\ref{FIG: stress/strain}c,f,i) suggest the ability to redistribute and relax accumulated stresses via plastic deformation.
Subsequent structural analysis reveals that key nm-scale mechanisms characterising out-of-plane loading are M--B layer ripples, sliding, and buckling, or nucleation of ripplocations and mechanical twins (Fig.~\ref{FIG: out-of-plane loading}).

Generally, [001] tensile strain first causes small layer ripples, nucleating uniformly throughout the lattice and resembling high-frequency small-amplitude, most often anti-phase waves (Fig.~\ref{FIG: out-of-plane loading}a1).
These facilitate work hardening.
Further [001] strain increase relaxes these ripples via M--A/B layer sliding (Fig.~\ref{FIG: out-of-plane loading}a2) and, subsequently, the layers start bucking similarly to low-frequency, most often in-phase waves (Fig.~\ref{FIG: out-of-plane loading}a3). 
As buckling becomes locally more pronounced, turning into irreversible kink boundaries, some layers delaminate (Fig.~\ref{FIG: out-of-plane loading}c), resulting in nanocrack growth approximately along (001) planes. 
This is consistent with {\it{ab initio}} MD tensile tests showing layer ripples, sliding, and delamination, followed by formation of approximate $(001)$ surfaces (Fig.~\ref{FIG: out-of-plane loading}a4).

Differences between individual MABs lie in their more-or-less-strong ability to nucleate anti-phase M--A layer ripples and retard kinking-induced failure via uniform buckling or ripplocations (Fig.~\ref{FIG: out-of-plane loading}b).
In line with atomic-scale {\it{ab initio}} MD, anti-phase layer ripples in nanoscale MLIP-MD simulations almost always nucleate along the [010] direction and are significantly less pronounced in 314 type MABs (Fig.~\ref{FIG: out-of-plane loading}b1), where they also live within shorter strain range (Fig.~\ref{FIG: out-of-plane loading}b2).
This may be intuited via higher strain/energetic costs of bending thicker ceramic-like layers of 314 type MABs.
Generally, Al seems to act like a lubricant facilitating movements of the stiffer M--B sublattice.
Ti$_2$AlB$_2$, Ta$_2$AlB$_2$, and WAlB exemplify MABs efficiently postponing the onset of  kinking-induced mechanical failure by gradual layer buckling. 
Contrarily in TiAlB, these buckles almost immediately turn into irreversible kink boundaries, causing an early fracture.

W$_2$AlB$_2$ and W$_3$AlB$_4$ do not exhibit anti-phase layer ripples.
Instead, the former deforms via ripplocations and twinning (Fig.~\ref{FIG: out-of-plane loading}c), while the latter forms stacking faults within Al layers (Fig.~\ref{FIG: out-of-plane loading}d).
These (and other) stackings faults are observed in {\it{ab initio}} MD tensile tests for W$_3$AlB$_4$ as well as for other MABs, e.g., Ta$_2$AlB$_2$ and W$_2$AlB$_2$ (note, however, the 2\% strain step, no Poisson's contraction and severe size constraints of {\it{ab initio}} MD). 
At the nanoscale, stacking changes appear as a local by-product of M--B/A layer sliding but only in W$_3$AlB$_4$, they form throughout nearly the entire supercell and live within fairly large strain range (Fig.~\ref{FIG: out-of-plane loading}b2).
M--B/A layer sliding---accompanied by stacking changes in Al layers---is an important mechanism also for W$_2$AlB$_2$ where it mediates changes in the twinning angle, namely its decrease, mirrored by increasing strain concentration and, ultimately, void opening.

\subsection{Temperature effects}
Constituting M--B layers of MABs, transition metal borides belong to ultra-high temperature ceramics (UHTCs) with outstanding high-temperature stability (melting points 2500--3500~K), hardness, resistance to corrosion, abrasive and erosive wear~\cite{magnuson2022review}.
Here we compare mechanical response of MABs at room and elevated temperature, represented by 1200~K (chosen in view of our validation against {\it{ab initio}} MD data, Tab.~\ref{TAB: aLat, Cij}, Fig.~\ref{FIG: validation}, and oxidation experiments for MABs~\cite{achenbach2019synthesis,lu2019crack} at temperatures relevant for cutting tool applications).

\begin{figure*}[h!t!]
    \centering
    \includegraphics[width=1.75\columnwidth]{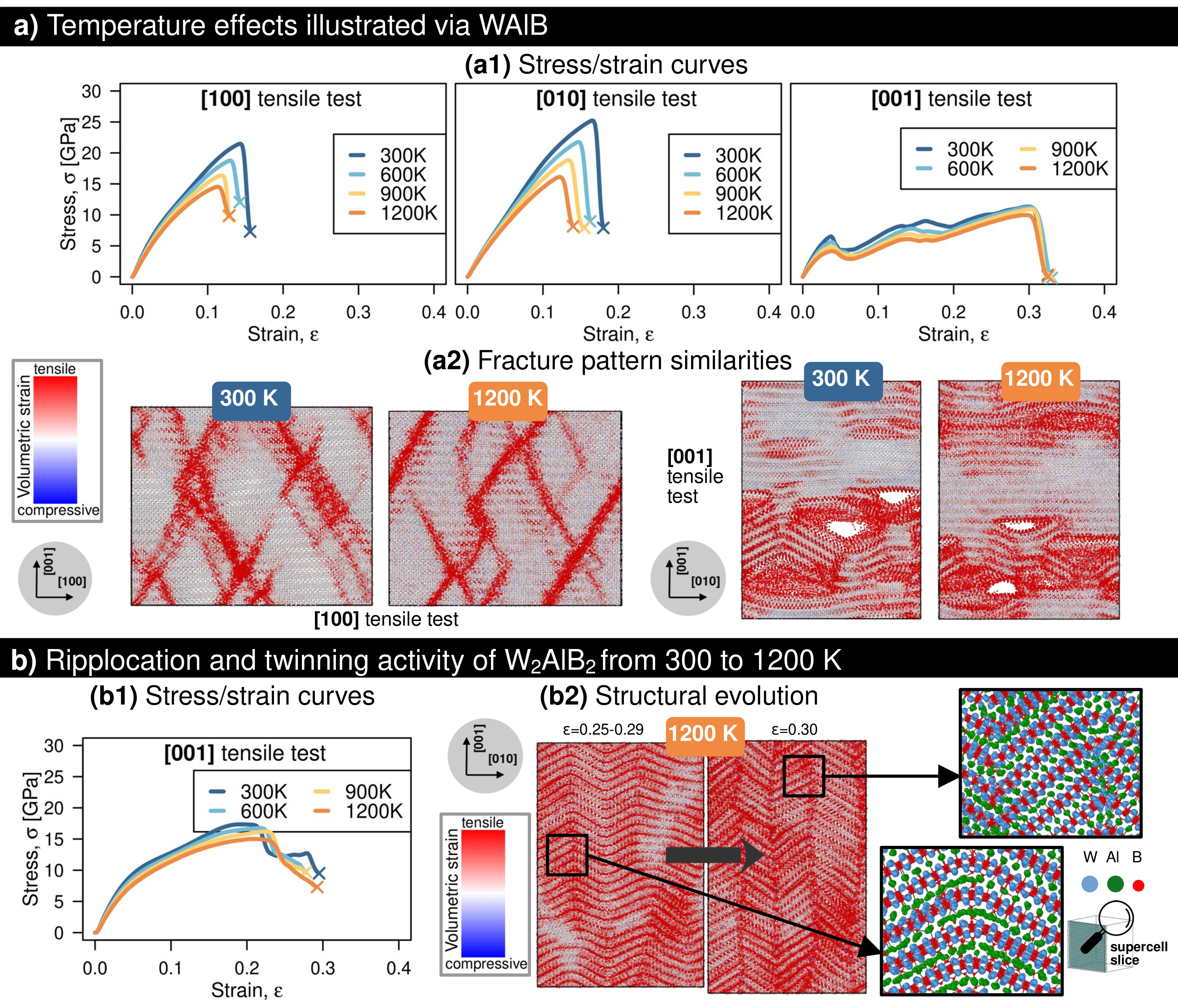}
    \caption{
    \small
    {\bf{High-temperature mechanical response of MABs derived from nanoscale MLIP-MD simulations}}.
    Computational details were consistent with previous tensile tests at 300~K (Fig.~\ref{FIG: stress/strain})
    (a) Typical $T$ effects in stress/strain curves (showing more significant decrease subject to in-plane loading) and deformation mechanisms (remaining qualitatively unchanged), as illustrated by WAlB.
    (b) Negligible changes of ripplocation and twinning abilities of W$_2$AlB$_2$.
    }
\label{FIG: T effects}
\end{figure*}

Besides expectable temperature-driven decline in theoretical tensile strength and toughness, mechanical properties (Tab.~\ref{TAB: T effects in mechanical properties}) and deformation mechanisms (Fig.~\ref{FIG: T effects}) at 1200~K remain consistent with those predicted at 300~K.
This includes work hardening subject to [001] tensile tests and ripplocation activity of W$_2$AlB$_2$.
Exceptions are TaAlB and TiAlB losing the ability of work hardening during [100] loading (TaAlB) and gaining the ability to ripplocate during [001] loading (TiAlB).
High-temperature mechanical response is generally characterised by larger atomic displacements, more frequent layer sliding, and (local) changes of the Al stacking sequence (similar to Fig.~\ref{FIG: out-of-plane loading}d).
It is also less strength-anisotropic, especially due to $\sigma_{\text{[100]}}$ and $\sigma_{\text{[010]}}$ decreasing generally faster, by $(27\pm5)\%$, compared with $(14\pm9)\%$ decrease of $\sigma_{\text{[001]}}$, where standard deviations reflect values for all MABs.

\begin{table*}[!htbp]
\centering
\footnotesize
\setlength{\tabcolsep}{4pt}
\caption{ \footnotesize
{\bf{High-temperature mechanical properties of the nine studied MABs, M$=$(Ti, Ta, W), A$=$Al, derived from nanoscale MLIP-MD simulations at 1200~K}}.
Computational details were consistent with previous tensile tests at 300~K (Fig.~\ref{FIG: stress/strain}) and the notation follows Tab.~\ref{TAB: mechanical properties}.
Specifically, $\sigma_{\text{[100]}}^{\text{max}}$ ($U_{\text{[100]}}^{\text{max}}$), $\sigma_{\text{[010]}}^{\text{max}}$ ($U_{\text{[010]}}^{\text{max}}$), and $\sigma_{\text{[001]}}^{\text{max}}$ ($U_{\text{[001]}}^{\text{max}}$) denote theoretical tensile strength (toughness) in the [100], [010], and [001] direction, respectively.
Percentages in brackets, $\textcolor{blue}{{[X\%]}}$, show differences from the corresponding  value at 300~K. 
} 
\label{TAB: T effects in mechanical properties}
  \begin{tabular}{ccccccccccccc}
    \hline
    \hline
    \multicolumn{1}{c}{{\bf{M}}} & \multicolumn{1}{|c}{\bf MAB} & \multicolumn{1}{|c}{\bf Number of} & \multicolumn{1}{|c|}{\bf Temp.~{\bf{[K]}}} & \multicolumn{4}{c|}{\bf Theoretical tensile strength [GPa]} & \multicolumn{4}{c}{\bf  Theoretical tensile toughness [GPa]} \\ 
     &  \multicolumn{1}{|c}{\bf phase} & \multicolumn{1}{|c}{\bf atoms } & \multicolumn{1}{|c|}{$T$} & $\sigma_{\text{[100]}}^{\text{max}}$ & $\sigma_{\text{[010]}}^{\text{max}}$ & $\sigma_{\text{[001]}}^{\text{max}}$ & \multicolumn{1}{c|}{ $\overline{\sigma}^{\text{max}}$ } & $U_{\text{[100]}}$ & $U_{\text{[010]}}$ & $U_{\text{[001]}}$ & $\overline{U}$ \\
     \hline
    {\bf{Ti}}& {\bf{222}} & 243,000 & 1200 & 15.5 & 16.3 & 6.2 & 12.7$\pm$5.6 & 1.11 & 1.89 & 0.75 & 1.2$\pm$0.6  \\ 
      &  & & & {\scriptsize\textcolor{blue}{$[-27\%]$}} & {\scriptsize\textcolor{blue}{$[-31\%]$}} & {\scriptsize\textcolor{blue}{$[-19\%]$}} & {\scriptsize\textcolor{blue}{$[-27\%]$}} & {\scriptsize\textcolor{blue}{$[-38\%]$}} & {\scriptsize\textcolor{blue}{$[-45\%]$}} & {\scriptsize\textcolor{blue}{$[-9\%]$}} & {\scriptsize\textcolor{blue}{$[-40\%]$}}  \\ 
    &{\bf{212}} & 286,650 & 1200 & 21.3 & 22.3 & 13.2 & 18.9$\pm$5.0 & 1.73 & 2.96 & 2.91 & 2.50$\pm$0.70 \\ 
    &  &  & & {\scriptsize\textcolor{blue}{$[-24\%]$}} & {\scriptsize\textcolor{blue}{$[-25\%]$}} & {\scriptsize\textcolor{blue}{$[-23\%]$}} & {\scriptsize\textcolor{blue}{$[-24\%]$}} & {\scriptsize\textcolor{blue}{$[-26\%]$}} & {\scriptsize\textcolor{blue}{$[-36\%]$}} & {\scriptsize\textcolor{blue}{$[-25\%]$}} & {\scriptsize\textcolor{blue}{$[-31\%]$}}  \\ 
    &{\bf{314}} & 294,914 & 1200 & 29.1 & 27.3 & 17.5 & 24.6$\pm$6.3 & 2.40 & 3.37 & 2.77 & 2.80$\pm$0.50 \\ 
     & & & & {\scriptsize\textcolor{blue}{$[-22\%]$}} & {\scriptsize\textcolor{blue}{$[-22\%]$}} & {\scriptsize\textcolor{blue}{$[-2\%]$}} & {\scriptsize\textcolor{blue}{$[-18\%]$}} & {\scriptsize\textcolor{blue}{$[-31\%]$}} & {\scriptsize\textcolor{blue}{$[-27\%]$}} & {\scriptsize\textcolor{blue}{$[-5\%]$}} & {\scriptsize\textcolor{blue}{$[-24\%]$}} \\ 
    \hline
    {\bf{Ta}} & {\bf{222}}  & 243,000 & 1200 & 12.0& 18.7 & 8.9 & 13.2$\pm$5 & 1.13 & 2.09 & 2.03 & 1.80$\pm$0.50 \\ 
     & &  & & {\scriptsize\textcolor{blue}{$[-30\%]$}} & {\scriptsize\textcolor{blue}{$[-30\%]$}} & {\scriptsize\textcolor{blue}{$[-7\%]$}} & {\scriptsize\textcolor{blue}{$[-26\%]$}} & {\scriptsize\textcolor{blue}{$[-65\%]$}} & {\scriptsize\textcolor{blue}{$[-50\%]$}} & {\scriptsize\textcolor{blue}{$[61\%]$}} & {\scriptsize\textcolor{blue}{$[-38\%]$}} \\ 
    & {\bf{212}} & 286,650 & 1200 & 13.9 & 21.7 & 13.2 & 16.2$\pm$4.7 & 2.00 & 2.50 & 2.95& 2.10$\pm$0.30 \\ 
     & & & & {\scriptsize\textcolor{blue}{$[-29\%]$}} & {\scriptsize\textcolor{blue}{$[-25\%]$}} & {\scriptsize\textcolor{blue}{$[-4\%]$}} & {\scriptsize\textcolor{blue}{$[-22\%]$}} & {\scriptsize\textcolor{blue}{$[-45\%]$}} & {\scriptsize\textcolor{blue}{$[-38\%]$}} & {\scriptsize\textcolor{blue}{$[-11\%]$}} & {\scriptsize\textcolor{blue}{$[-36\%]$}} \\ 
    & {\bf{314}} & 294,914 & 1200 & 21.2 & 26.5 & 14.2 & 20.7$\pm$6.2 & 2.08 & 3.23 & 2.00 & 2.40$\pm$0.70 \\ 
     & & &  & {\scriptsize\textcolor{blue}{$[-17\%]$}} & {\scriptsize\textcolor{blue}{$[-24\%]$}} & {\scriptsize\textcolor{blue}{$[-9\%]$}} & {\scriptsize\textcolor{blue}{$[-18\%]$}} & {\scriptsize\textcolor{blue}{$[-27\%]$}} & {\scriptsize\textcolor{blue}{$[-45\%]$}} & {\scriptsize\textcolor{blue}{$[+14\%]$}} & {\scriptsize\textcolor{blue}{$[-31\%]$}} \\ 
    \hline
    {\bf{W}} & {\bf{222}} & 243,000 & 1200 & 14.8 & 16.3 & 10.2 & 13.8$\pm$3.2 & 1.05 & 1.16 & 1.76 & 1.30$\pm$0.40 \\ 
    &  &  & & {\scriptsize\textcolor{blue}{$[-32\%]$}} & {\scriptsize\textcolor{blue}{$[-36\%]$}} & {\scriptsize\textcolor{blue}{$[-12\%]$}} & {\scriptsize\textcolor{blue}{$[-30\%]$}} & {\scriptsize\textcolor{blue}{$^[-45\%]$}} & {\scriptsize\textcolor{blue}{$[-54\%]$}} & {\scriptsize\textcolor{blue}{$[-26\%]$}} & {\scriptsize\textcolor{blue}{$[-43\%]$}} \\ 
    & {\bf{212}} & 286,650 & 1200 & 20.2 & 22.4 & 15.0 & 19.2$\pm$3.8 & 1.93 & 2.10 & 2.19 & 2.10$\pm$0.10 \\ 
     &  &  & & {\scriptsize\textcolor{blue}{$[-26\%]$}} & {\scriptsize\textcolor{blue}{$[-30\%]$}} & {\scriptsize\textcolor{blue}{$[-14\%]$}} & {\scriptsize\textcolor{blue}{$[-25\%]$}} & {\scriptsize\textcolor{blue}{$[-36\%]$}} & {\scriptsize\textcolor{blue}{$[-45\%]$}} & {\scriptsize\textcolor{blue}{$[-1\%]$}} & {\scriptsize\textcolor{blue}{$[-30\%]$}}  \\ 
    & {\bf{314}} & 294,914 & 1200 & 21.1 & 23.1 & 15.5 & 19.9$\pm$3.9 & 1.94 & 2.02 & 2.97 & 2.30$\pm$0.60 \\ 
    & & & & {\scriptsize\textcolor{blue}{$[-28\%]$}} & {\scriptsize\textcolor{blue}{$[-21\%]$}} & {\scriptsize\textcolor{blue}{$[-33\%]$}} & {\scriptsize\textcolor{blue}{$[-37\%]$}} & {\scriptsize\textcolor{blue}{$[-36\%]$}} & {\scriptsize\textcolor{blue}{$[-20\%]$}} & {\scriptsize\textcolor{blue}{$[-10\%]$}} & {\scriptsize\textcolor{blue}{$[-21\%]$}} \\     
    \hline
    \hline
\end{tabular}
\end{table*}

A slight decline of [001] tensile strength and toughness is predicted for Ti$_3$AlB$_4$, in which $\sigma_{\text{[001]}}$ and $U_{\text{[001]}}$ decrease by only a 2 and 5~\%, respectively, as $T$ rises from 300 to 1200~K (Tab.~\ref{TAB: T effects in mechanical properties}).
Other examples of small temperature effects along [001] direction are Ta$_2$AlB$_2$, Ta$_3$AlB$_4$, and W$_2$AlB$_2$ yielding a $\sigma_{\text{[001]}}$ and $U_{\text{[001]}}$ decrease of 4--14\% and 1--14~\%, respectively. 
The experimentally known WAlB exhibits a fairly large drop of both strength and toughness, although, again, less pronounced normal to basal planes (Fig.~\ref{FIG: T effects}a): $\sigma_{\text{[001]}}$ and $U_{\text{[001]}}$ decline by 12 and 26~\%, respectively, while $\sigma_{\text{[100]}}$ and $U_{\text{[100]}}$ decline by 32 and 45~\%, respectively.

Temperature effects on ripplocation and twinning activities of W$_2$AlB$_2$ (Fig.~\ref{FIG: T effects}b) are minimal, mirrored by $U_{\text{[001]}}$ decrease of 1\% (Tab.~\ref{TAB: T effects in mechanical properties}).
Additionally to simulations at $T\in\{300,600,900,1200\}$~K, we tested various supercell sizes and included a free surface, all pointing towards nucleation of ripplocations and twins during [001] tensile loading. 
TiAlB presents similar ripplocation-based response to [001] strain within temperatures range 600--1200~K.
This phase, however, may be difficult to synthesise (with $E_f$ about 0.24~eV/at. above that of Ti$_3$AlB$_4$~\cite{koutna2024phase}) and it is also overshadowed by W$_2$AlB$_2$ in terms of theoretical tensile strength and toughness in all directions. 

\subsection{Design implications}
Among our most interesting predictions is the ability of W$_2$AlB$_2$---with formation energy 0.02~eV/at. above that of the experimentally known WAlB~\cite{koutna2024phase}---to form ripplocations, followed by nucleation of mechanical twins.
Both ripplocations and twins are activated in a wide temperature range and are likely responsible for only minor toughness decrease from 300 up to 1200~K.

Generally, twins are known to enhance the hardness, toughness, thermal stability, as well as wear resistance of materials, however, their formation in ceramics is typically prohibited by high stacking-fault energies~\cite{chen2023cross,huang2025harvesting}.
The modulated ceramic-like/metallic-like bonding character of MABs may provide more favourable conditions for stacking faults, as indicated by both {\it{ab initio}} MD as well as MLIP-MD simulations for W$_2$AlB$_2$ but also for other MABs, e.g., TiAlB, Ta$_2$AlB$_2$, W$_3$AlB$_3$. 
Intuitively, low stacking fault energies may not be the only decisive factor for twinning in MABs.
We speculate that twinning is related to the ability to ripplocate, based on simulations for W$_2$AlB$_2$ and TiAlB (above 300~K).
Instead of highly-symmetric twin boundaries, other MABs nucleate kink boundaries, preceded by layer buckling which, however, seems too short and has too low angles to be called ripplocations. 

How likely is detecting ripplocations or twins in W$_2$AlB$_2$ under real lab-scale conditions?
In MAX phases---with atomically-laminated structures similar to MABs---ripplocations and twins have been reported during indentation tests comprising basal-plane compression ~\cite{plummer2021origin,tromas2022nanoindentation,parent2024atomic}.  
Although our loading is tensile, the $\approx{7}$~\% in-plane Poisson compression at the onset of ripplocation activity could mimic regions near the indenter tip.
An important question for follow-up research concerns interactions between ripplocations and typical growth defects (e.g., vacancies, dislocations, grain boundaries) in W$_2$AlB$_2$. 
In this regard, little is known for any material (see, e.g., studies of graphite showing that ripplocations migrate towards vacancies and annihilate~\cite{gruber2020characterization}).

Besides W$_2$AlB$_2$, other promising MAB phase candidates include Ta$_2$AlB$_2$ and W$_3$AlB$_4$---providing a suitable basis for high stacking fault density, thus, fine-tuning plasticity---or Ti$_3$AlB$_4$ providing the highest theoretical tensile strength.
In view of [001]-tensile-strain-induced work hardening (predicted for all MABs) as well as ripplocation and twinning activities (W$_2$AlB$_2$, TiAlB), it may be advantageous growing MABs with basal planes normal to the substrate, as e.g. Cr$_2$AlB$_2$~\cite{berastegui2020magnetron}.
During nanoindentation, MABs with such preferential orientation could exhibit in-plane tensile stresses similar to the simulated ones, facilitating work hardening and superior fracture resistance.

\section{Conclusions}
Tensile response of the group 4--6 transition metal based single-crystal MAB phases---with $\text{M}=(\text{Ti}$, Ta, W), $\text{A}=\text{Al}$, and the 222, 212, and 314 type phases---was screened by nanoscale machine-learning powered molecular dynamics simulations, both at 300~K as  well as at elevated temperatures.
Key deformation/failure mechanisms included (i) slipping, the easiest along the $\{110\} \langle001\rangle$ slip system, and shear-banding subject to in-plane tensile loads; (ii) layer buckling---mirrored by strain hardening---and kinking subject to loads normal to basal planes; (iii) ripplocations and twins nucleating in W$_2$AlB$_2$.

{\it{Ab initio}} data set for active learning of our MLIPs was generated semi-automatically, with little material-specific knowledge; by simulating series of atomic-scale mechanical tests ($<{10}^3$-atom supercells, finite $T$) triggering growth of various defective environments (e.g., stacking faults, buckled or mixed layers, voids, surfaces).
This strategy is likely transferable to other MABs. 
Validation concerned statistical errors of atomic properties evaluated against $\approx{3\cdot10^6}$ DFT/{\it{ab initio}} MD configurations, predictions of observables, and extrapolation grade analysis.

Ti$_3$AlB$_4$ exhibited the highest theoretical tensile strength ($\approx{45}$\% of that for a typical hard ceramic, TiB$_2$~\cite{lin2024machine}), whereas Ta$_2$AlB$_2$, W$_2$AlB$_2$, Ta$_3$AlB$_4$, W$_3$AlB$_4$ showed outstanding toughness. 
Temperature effects were generally anisotropic: causing faster deterioration of mechanical properties in-plane, compared with mild effects along [001].
We suggest that Ta- and W-based MABs with the 212 or 314 phase are especially promising for experimental investigations including micromechanical testing and TEM studies.
Ta$_2$AlB$_2$ and W$_3$AlB$_4$ are suitable for tuning transformation plasticity via stacking fault formation, while W$_2$AlB$_2$ allows for toughness enhancement via strain-activated ripplocations.

The---nearly temperature-independent---ripplocation and twinning ability of W$_2$AlB$_2$ seems unique and may stem from its particular electronic structure, easy Al layer sliding, and energy landscape of certain stacking faults.
These are proposed as important directions of follow-up research, together with interactions between ripplocations and typical growth defects, as well as the response of MABs to mechanical loads in the presence of a pre-crack. 
Enhancing the transferability of our MLIPs to such simulations will require active learning on under-represented environments, e.g., highly compressed, off-stoichiommetric, and featuring planar defects. 

We illustrated that MLIPs for complex material systems can be trained in semi-automatic fashion and used to predict temperature evolution of deformation-related descriptors, inherent strengthening mechanisms as well as nucleation of plasticity-enhancing defects at scales accessible to experiment.
Considering phase stability trends and other properties of MABs being mostly driven by the M element~\cite{koutna2024phase}, our training sets may accelerate MLIP training for MABs of all group 4--6 transition metals and various A species.

\section*{CRediT authorship contribution statement}
{\bf{NK}}: Conceptualisation, Methodology, Investigation, Data curation, Visualisation, Writing – original draft. 
{\bf{SL}}: Data curation, Validation, Writing – review \& editing. 
{\bf{LH}}: Resources, Writing – review \& editing. 
{\bf{DGS}}: Conceptualisation, Methodology, Resources, Writing – review \& editing. 
{\bf{PHM}}: Resources, Writing – review \& editing. 

\section*{Declaration of Competing Interests}
The authors declare no competing interests.

\section*{Data Availability}
The data presented in this study are available from the corresponding author upon reasonable request.

\section*{Acknowledgements}
{\bf{NK}} and {\bf{PHM}} acknowledge the Austrian Science Fund, FWF, 10.55776/PAT4425523. 
{\bf{LH}} acknowledges financial support from the Swedish Government Strategic Research Area in Materials Science on Functional Materials at Link\"{o}ping University SFO-Mat-LiU No. 2009 00971 and the support from Knut and Alice Wallenberg Foundation Scholar Grant KAW2019.0290.
{\bf{DGS}} acknowledges financial support from the Swedish Research Council (VR) through Grant Nº VR-2021-04426 and the Competence Center Functional Nanoscale Materials (FunMat-II) (Vinnova Grant No. 2022-03071).
The computations handling were enabled by resources provided by the National Academic Infrastructure for Supercomputing in Sweden (NAISS) and the Swedish National Infrastructure for Computing (SNIC) at the National Supercomputer Center (NSC) partially funded by the Swedish Research Council through grant agreements no. 2022-06725 and no.~2018-05973, as well as by the Vienna Scientific Cluster (VSC) in Austria. 
The authors acknowledge TU Wien Bibliothek for financial support through its Open Access Funding Programme.

\bibliography{references.bib}

\end{document}